\documentclass[aps,superscriptaddress,preprint]{revtex4-1}

\usepackage{threeparttable}
\usepackage[dvips]{graphicx}
\usepackage{color}
\usepackage{natbib}

\newcommand{\hl}[1]{#1}

\begin{document}

%\onecolumn
%\firstpage{1}

\title[Mobility of lanthanide and actinide cations]{Mobility of the singly-charged lanthanide and actinide cations: trends and perspectives}

%\affiliation{Skolkovo Institute of Science and Technology, Skolkovo Innovation Center, Moscow  121205,  Russia}

\author{Giorgio Visentin}
\affiliation{Skolkovo Institute of Science and Technology, Moscow,  Russia}
\author{Mustapha Laatiaoui}
\affiliation{Johannes Gutenberg-Universit{\"a}t, Mainz, Germany}
\affiliation{Helmholtz-Institut Mainz, Mainz, Germany}
\author{Larry A. Viehland}
\affiliation{Science Department, Chatham University, Pittsburgh, Pennsylvania, USA}
\author{Alexei A. Buchachenko}\email{a.buchachenko@skoltech.ru}
\affiliation{Skolkovo Institute of Science and Technology, Moscow,  Russia}
\affiliation{Institute of Problems of Chemical Physics RAS, Chernogolovka, Russia}

%\author[\firstAuthorLast ]{\Authors} %This field will be automatically populated
%\address{} %This field will be automatically populated
%\correspondance{} %This field will be automatically populated

%\extraAuth{}% If there are more than 1 corresponding author, comment this line and uncomment the next one.
%\extraAuth{corresponding Author2 \\ Laboratory X2, Institute X2, Department X2, Organization X2, Street X2, City X2 , State XX2 (only USA, Canada and Australia), Zip Code2, X2 Country X2, email2@uni2.edu}

\begin{abstract}
%%% Leave the Abstract empty if your article does not require one, please see the Summary Table for full details.
%\section{}
The current status of gaseous transport studies of the singly-charged lanthanide and actinide ions is reviewed in light of potential applications to superheavy ions. The measurements and calculations for the mobility of lanthanide ions in He and Ar agree well, and they are remarkably sensitive to the electronic configuration of the ion, namely, whether the outer electronic shells are 6s, 5d6s or 6s$^2$. The previous theoretical work is extended here to ions of the actinide family with zero electron orbital momentum: Ac$^+$ (7s$^2$, $^1$S), Am$^+$ (5f$^7$7s $^9$S$^\circ$), Cm$^+$ (5f$^7$7s$^2$ $^8$S$^\circ$), No$^+$ (5f$^{14}$7s $^2$S) and Lr$^+$ (5f$^{14}$7s$^2$ $^1$S). The calculations reveal large systematic differences in the mobilities of the 7s and 7s$^2$ groups of ions and other similarities with their lanthanide analogs. The correlation of ion-neutral interaction potentials and mobility variations with spatial parameters of the electron distributions in the bare ions is explored through the ionic radii concept. While the qualitative trends found for interaction potentials and mobilities render them appealing for superheavy ion research, lack of experimental data and limitations of the scalar relativistic {\it ab initio} approaches in use make further efforts necessary to bring the transport measurements into the inventory of techniques operating in ``one atom at a time'' mode.

\vspace{1cm}
Keywords: ion mobility, interaction potential, lanthanides, actinides, electronic configuration, superheavy ions %All article types: you may provide up to 8 keywords; at least 5 are mandatory.

\end{abstract}

\maketitle

\section{Introduction}
\label{sec:Intro}

While celebrating 1869 as the year of the Periodic Table's discovery, one may also recall other important milestones of its shaping towards the present form~\citep{Karol2016b,Karol2016a}. The last element found in nature was francium $Z=87$ in 1939~\citep{Perey1939}, although a few more have been confirmed after being produced artificially. The synthetic era started in 1937 with technetium $Z=43$~\citep{Perrier1947}. The transuranium elements up to fermium $Z=100$, discovered in 1952~\citep{Ghiorso1955b}, are produced in nuclear reactors by neutron capture reactions. About the same time, mendelevium $Z = 101$ was synthesized~\citep{Ghiorso1955a} by a new recoil technique in ``one atom at a time'' mode. This technique has opened the modern era of heavy ion fusion synthesis that is still being used in high power accelerators~\citep{Turler2013, Haba2019}.

Although recent discoveries have been driven by physical methods, it is essentially chemistry that fit them into the Periodic Table. Even the actinides had not found their proper placement until the chemical analysis of neptunium $Z = 93$ and plutonium $Z = 94$ in the 1940's~\citep{Seaborg1945, Seaborg1990}. Since then, the chemical isolation of an element marks its discovery~\citep{Wallmann1959}. Sophisticated techniques of production, isolation and characterization of simple chemical compounds in both gas and liquid phases are in use~\citep{Turler2013, SchadelBook2014, Oganessian2016, Eichler2017, Eichler2019, Duellmann2019} for superheavy elements to determine their volatility, adsorption enthalpies and bonding parameters.

Information on the electronic structure and properties of bare heavy atoms and ions is no less valuable. In particular, spectroscopic data enables firm assignments of ground state configurations, irrespective of the chemical behavior. In addition, it provides fingerprint spectral lines for use in the search for heavy and superheavy elements in the universe~\citep{TerAkopian2015} and benchmark data for {\it ab initio} methods of atomic and nuclear structure theory~\citep{Pershina1996, Pyykko2012, Eliav2015, Pyykko2016, LiuBook2017, Dzuba2017, Giuliani2019}. The recent review by~\cite{Backe2015} relates the progress in spectroscopic measurements to the use of ion or buffer gas traps to collect a few atomic species emerging one by one from a recoil separator. It acknowledges that ``quite good spectroscopic information is available up to the element einsteinium ($Z = 99$)... up to the year 2003''. Since then, the bound has been gradually pushed upward~\citep{Sewtz2003,Laatiaoui2016,Chhetri2018} to nobelium ($Z = 102$) owing to resonance ionization spectroscopy of the neutral atoms inside buffer gas cells.

The extension of these technique to heavier elements is certainly challenging, mostly due to decreasing production yield with increasing atomic number. Classical methods based on fluorescence detection suffer from low sensitivity, which renders them incompatible with one atom at a time experiments~\citep{Campbell2016}. Not surprisingly, studies of the gaseous transport properties are currently being considered as prospective means for probing the superheavy ions~\citep{Rickert2020}, not least for their compatibility with in-flight separators that provide recoil ions~\citep{Backe2015}.

From many measurements across the Periodic Table, gaseous ion mobility is known to be sensitive to the electronic configuration of open-shell ions~\citep{Kemper1991, Bowers1993, Taylor1999, Iceman2007, Ibrahim2008, Manard2016b, Manard2016a}. It is a fundamental property of an ion that defines, macroscopically, the rate of its steady-state drift through a neutral buffer gas and reflects its microscopic interactions with the buffer-gas particles~\citep{MasonBook1988, LarryBook2018}. In a sense, characterization of an ion through its gas-phase interaction with other species is equivalent to chemical characterization by chromatography. By choosing monoatomic inert gases as the buffers, one reduces the complexity of covalent chemical bonding to the (relative) simplicity of the physical ion-atom polarization forces.

The theory of intermolecular forces tells us that the properties of a weakly bound dimer can be reliably described by the properties of the constituting monomers~\citep{StoneBook, KaplanBook}. Thus, ion-atom interaction potentials are very sensitive to the electronic structure of an ion, to its electronic configuration, electronic state symmetry, electric momenta, and static and dynamic polarizabilities. Exemplary confirmation of this for the main-group and transition-metal ions has been provided by~\cite{Breckenridge2002} and~\cite{WrightBreckenridge2010}. Ion mobility inherits this sensitivity. The field-induced drift discrimination of the ions in ground and excited electronic states~\citep{Kemper1991, Bowers1993, Taylor1999, Iceman2007, Ibrahim2008, Manard2016b, Manard2016a}, known as the electronic-state chromatography effect, is a direct consequence of the mobility variation with electronic configuration. It has been proposed recently~\citep{LRC2019} that this effect can be used for spectroscopic investigation of heavy and superheavy ions.

Measurements of ion mobility (equivalently, the drift time through a fixed distance) are indeed compatible with the separation and buffer gas trapping techniques. They are well controlled by operating temperature, pressure and external field strength. Potentially, they can enrich our knowledge of electronic structure of ions produced in one atom at a time mode.

The present paper addresses the current state-of-the-art in the studies of gaseous transport of singly-charged lanthanide and actinide ions. Though far from being complete, experimental and theoretical data for the lanthanide ions still permit us to analyze the relation between the electronic structure of an ion and its mobility determined by the {\it ab initio} ion-atom interaction potential. In particular, mobility trends for distinct electronic configurations and effective sizes of an ion are established. To step into the actinide period, we extend the \hl{scalar relativistic} {\it ab initio} approaches tested for lanthanides to compute ion-atom interaction potentials for selected actinide ions. We show that the trends found for the lanthanides largely persist for the actinide family and thus can underlie experimental exploration of their transport and, in turn, electronic structure properties. This also sheds the light on potential use of transport properties for exploration of superheavy ions.

In sec.~\ref{sec:ionmob} we briefly review the theoretical concepts and computational methods of ion mobility in rare gases. Sec.~\ref{sec:la} presents the review and analysis of the lanthanide results, while ions of the actinide family are discussed in Sec.~\ref{sec:ac}. Conclusions and outlook follow.

\section{Ion mobility and interaction potentials}
\label{sec:ionmob}

Experimental techniques, general theoretical concepts and computational approaches relevant to gaseous ion transport are described in detail in two monographs by~\cite{MasonBook1988} and by~\cite{LarryBook2018}. The macroscopic definition of the mobility, $K$, for trace amounts of drifting ions is given by the equation
\begin{equation}
\label{eq:K}
{\mathbf v}_d = K\mathbf{E},
\end{equation}
where the vector, ${\mathbf v}_d$, is the ion drift velocity and $\mathbf{E}$ is the electric field vector. Throughout this paper, only the monoatomic rare gases He and Ar \hl{(collectively, RG)} are considered as the buffer gases. \hl{The ion mobility is deduced from the measured arrival time distribution of the ions drifting through the tube of length $l$. In particular, the mean drift time $t_d$ is}
\begin{equation}
\label{eq:td}
t_d = l/KE.
\end{equation}

It is convenient to consider the standard mobility, $K_0$, by the equation
\begin{equation}
\label{eq:K0}
K_0 = n_0K/N_0,
\end{equation}
where $n_0$ and $N_0=2.6867805$ m$^{-3}$ are the buffer gas number density and the Loschmidt number, respectively. The standard mobility depends on the reduced electric field strength, $E/n_0$, and the temperature of the gas, $T_0$.

From a rigorous theoretical standpoint, the ion mobility is a transport coefficient determined by the solution of the Boltzmann equation, which accounts for anisotropic diffusion and equilibration of the dragging electrostatic force by the momentum transfer that determines the stationary velocity of an ion through the buffer gas. The Boltzmann equation is parameterized by collision integrals, which are expressed through the binary collision cross sections~\citep{MasonBook1988,LarryBook2018}. The cross sections are, in turn, fully determined by the ion-atom interaction potential(s). Vice versa, knowledge of the zero-field mobility over a reasonably wide range of $E/n_0$ or $T_0$ is enough for direct reconstruction of the interaction potential~\citep{Viehland1983,Viehland1976,MasonBook1988}.

The Gram-Charlier expansion of the ion distribution function provides the most sophisticated approach for solving the Boltzmann equation for atomic ions drifting in atomic gases~\citep{ViehlandGC1994,LarryBook2018}. Its accuracy has been shown to be limited solely by the accuracy of the underlying ion-atom potential~\citep{Viehland2012,Viehland2017}. The Gram-Charlier method is used for all mobility calculations considered in this paper.  The results of these calculations have been placed in the on-line database~\citep{LXCAT} within the LXCat project, that already has about 5000 tables of theoretical and experimental results.

In the low-field limit, which is the only situation considered here, $K_0$ has only a slight dependence on the gas temperature, as indicated by writing it as $K_0(T_0)$. The Gram-Charlier theory reduces \hl{in this situation} to the one-temperature theory~\citep{MasonBook1988,LarryBook2018} and \hl{the so-called zero-field mobility}, $K_0(T_0)$, obeys the fundamental low-field ion mobility equation~\citep{LarryBook2018}, which contains the momentum-transfer collision integral, $\bar{\Omega}^{(1,1)}(T_0)$. {According to~\cite{MasonBook1988} and~\cite{LarryBook2018}, this equation is
\begin{equation}
\label{eq:Langevin}
K_0(T_0) = \left(\frac{2\pi}{\mu_0k_BT_0}\right)^{1/2}\frac{3q}{16N_0}\frac{1+\alpha_c(T_0)}{\bar{\Omega}^{(1,1)}(T_0)},
\end{equation}
where $\mu_0$ is the reduced mass of the ion-atom system, $k_B$ is the Boltzmann constant, $q$ is the ion charge (always +1 in electron charge units here), and $\alpha_c(T_0)$ is a temperature-dependent correction term that is small enough to be neglected for heavy ions~\citep{LarryBook2018}.  Note that $\bar{\Omega}^{(1,1)}(T_0)$ has the standard definition~\citep{Hirschfelder1954} as the temperature average of the energy-dependent momentum-transfer cross section. Throughout this paper, the classical-mechanical cross sections were computed using the program PC~\citep{PC2010}.

A complication arises when an ion has an open-shell electronic structure, as is the case for the majority of singly-charged lanthanides and actinides. Non-zero electronic orbital angular momentum, $L$, makes the ion-atom interaction anisotropic~\citep{AquilantiGrossi1980,Krems2004}. The ion-atom collisions controlling the ion transport may involve multiple underlying interaction potentials and the respective cross sections depend on $\Lambda$, the projection of $L$ onto the collision axis. If, in addition, an ion bears non-vanishing electronic spin, $S$, vectorial spin-orbit (SO) interaction couples $L$ and $S$ into the total electronic angular momentum, $J$. The interaction remains anisotropic in $\Omega$, the projection \hl{of $J$} onto the collision axis, if $J \ge 1$. Moreover, if the SO splitting is small, inelastic fine-structure transitions can affect the transport at elevated $T_0$.

As the present paper primarily explores the relation between the ion electronic structure and the ion mobility, through the ion-atom interaction potential, we will mostly consider scalar-relativistic approaches. Vectorial SO coupling can be used in subsequent work for accurate comparisons with experimental data.

Consideration of interaction anisotropy gives rise to some ambiguity. A transparent one-to-one relation between the interaction potential and transport properties holds within the so-called ``isotropic scalar relativistic'' (ISR) approximation that was first introduced by~\cite{Aquilanti1989} for diffusion of neutral atoms. It assumes that the collisions changing $\Lambda$  are very efficient, so that an atom ``feels'' an ion through the average isotropic potential $V_0$. \hl{For ions in the states of D symmetry ($L=2$), like Gd$^+$ (4f$^7$5d6s, $^{10}$D$^\circ$) and metastable Lu$^+$ (4f$^{14}$5d6s, $^3$D) of relevance here, the isotropic potential has the form}
\begin{equation}
\label{eq:v0}
V_0(R) = [V_\Sigma(R) + 2V_\Pi(R) + 2V_\Delta(R)]/5,
\end{equation}
where $\Sigma$, $\Pi$ and $\Delta$ correspond to projections $|\Lambda| = 0, 1$ and 2, respectively, and $R$ is the ion-atom internuclear distance. Alas, the ISR approximation can be rather poor~\citep{IJMS2}. More accurate is the ``anisotropic'' approximation (ASR), which assumes the conservation of $\Lambda$  during each ion-atom collision. The ASR implies that the transport cross sections should be computed for each $V_\Lambda$ potential separately and then averaged with the same degeneracy factors as appeared in eq. (\ref{eq:v0}).

The sensitivity of the ion mobility to the interaction potential is well known~\citep{MasonBook1988,LarryBook2018}. Extensive comparisons by~\cite{Viehland2017} for ions lighter than caesium $Z = 55$ indicates that the potentials calculated using an accurate \hl{single-reference} {\it ab initio} technique, such as the CCSD(T) (coupled cluster with singles, doubles and noniterative triples) method, normally provide the zero-field mobilities accurate within 0.05\%. By contrast, multireference methods of the configuration interaction type (like MRCI, multireference configuration interaction) are not well suited for interaction potentials involving heavy ions. Accounting for the static electron correlation in a bare ion requires long expansions over configurations with multiple high-angular momentum shell occupancies, while the recovery of the dynamic correlation necessary to reproduce the polarization forces makes the problem intractable. As a result, {\it ab initio} interaction potential calculations fitting the accuracy required for transport properties are presently possible only for ions whose electron configurations are well described in the single reference approximation. This limits the variety of ions studied using the CCDS(T) method and considered below. \hl{Another concern is the strong relativistic effects inherent to heavy ions. Ion-atom interactions predominantly depend on the density of outermost electrons and could be less sensitive to relativity than, say, electronic energy levels or chemical bonding. Indeed, as we show below, scalar-relativistic effective core potentials for lanthanide ions permit one to reproduce the measured mobility quantitatively. The same level of accuracy cannot be guaranteed for actinide ions and no direct comparison between the measured and calculated mobilities is currently possible. However, we believe that scalar relativistic CCSD(T) method is still able to capture qualitative trends in interaction potential and mobility variations along the family, while the applications of the more elaborate relativistic method should be reserved for quantitative analysis to come.}

Throughout this paper, we will use the following notations for ions. Complete specification for Eu$^+$, for instance, $_{63}^{151}$Eu$^+$(4f$^7$6s, $^9$S$^\circ$), includes the nuclear charge $Z$, the number of neutrons in the nuclei, the electronic configuration of outer shells and the term symbol in the scalar relativistic approximation. Particular isotopes are specified mostly for measured or calculated mobility data. \hl{In transparent cases, some of these symbols will be omitted. When the SO splitting is considered explicitly, the $J$ subscript is added to the term symbol.}

\section{Lanthanide ions}
\label{sec:la}

\subsection{Overview}
\label{ssec:over}

Lanthanide ions provide a useful test case for assessments of the theoretical approaches to heavy ion mobility and analysis of information about an ion's electronic structure that can be derived from the limited measurements. The first relevant experimental study performed by~\cite{Mustapha2012} provided the zero-field mobilities in Ar at 300 K for the $_{63}^{151}$Eu$^+$ (4f$^7$6s, $^9$S$^\circ$), $_{64}^{156}$Gd$^+$ (4f$^7$5d6s, $^{10}$D$^\circ$),  $_{65}^{159}$Tb$^+$ (4f$^9$6s, $^7$H$^\circ$), $_{67}^{165}$Ho$^+$ (4f$^{11}$6s, $^5$I$^\circ$),  $_{68}^{168}$Er$^+$ (4f$^{12}$6s, $^4$H) and  $_{70}^{174}$Yb$^+$ (4f$^{14}$6s, $^2$S) ions. Shortly after, two of us reported the {\it ab initio} CCSD(T) interaction potentials and transport properties for the ground S-state ions Eu$^+$, Yb$^+$  and $_{71}^{175}$Lu$^+$(4f$^{14}$6s$^2$, $^1$S) in He, Ne, Ar, Kr and Xe for wide ranges of $T_0$ and $E/n_0$ (data available from the LXCat database~\citep{LXCAT}). For the Gd$^+$ ion in the rare gases, a combination of the CCSD(T) and MRCI methods was applied together with an asymptotic model for SO coupling. Simultaneously, the Eu$^+$ (4f$^7$6s) interactions with the rare gases were calculated by~\cite{TimEd} using the CCSD(T) method combined with the large-core effective core potentials (ECP), see also~\citep{JCP52}. Manard and Kemper measured the zero-field mobilities in He at 295 K, first for the same four ions~(\citeyear{Manard2017}) and then~(\citeyear{Manard2017all}) for the rest of the lanthanide family from $_{58}^{140}$Ce$^+$ (4f5d$^2$, $^4$H$^\circ$) to $_{71}^{175}$Lu$^+$ (4f$^{14}$6s$^2$) except for $_{61}$Pm$^+$ (4f$^5$6s, $^7$H$^\circ$). More sophisticated {\it ab initio} calculations have allowed us to bring the Gd$^+$ ion mobilities in He and Ar into agreement with the measurements~\citep{IJMS2} and to evaluate the interaction potentials for the metastable $_{71}^{175}$Lu$^+$ (4f$^{14}$5d6s, $^3$D), as presented here.

\begin{figure}[hbt!]
\includegraphics[width=1.0\linewidth]{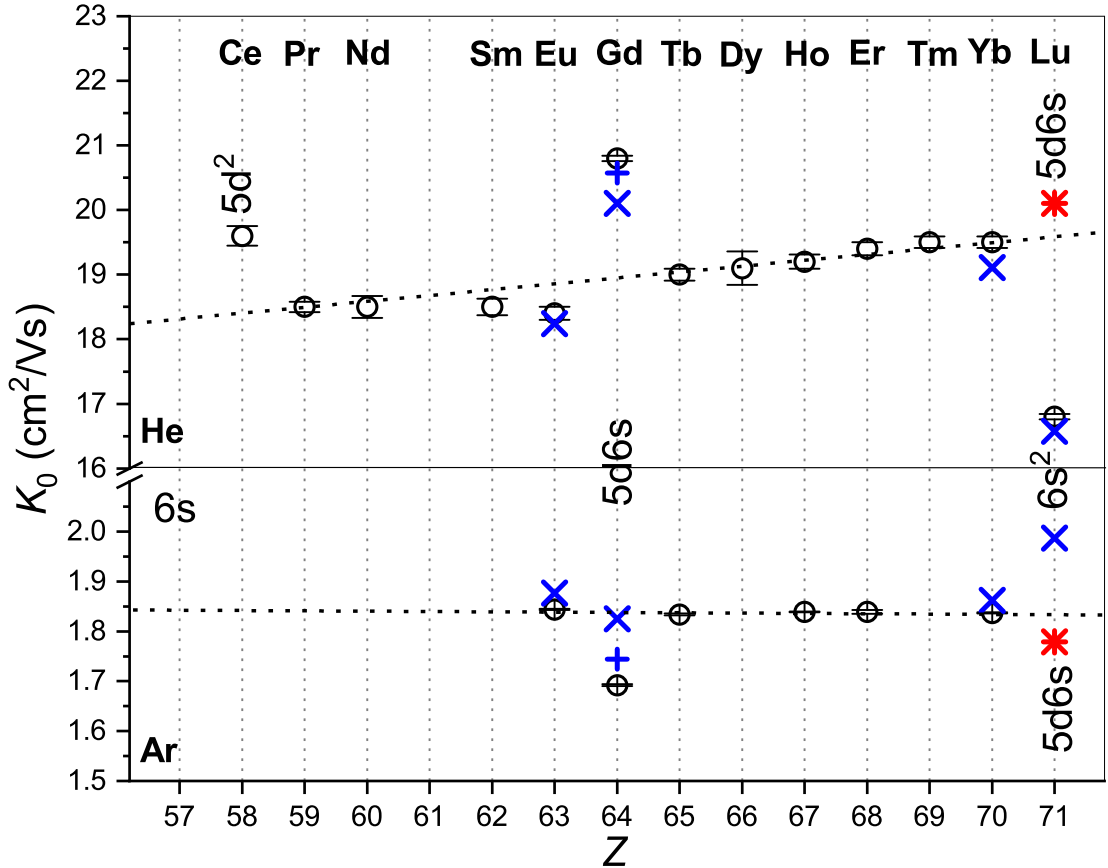}
\caption{
Comparison of the room-temperature zero-field mobilities of the lanthanide ions in He (top panel) and Ar (bottom panel). Circles indicate experimental data by~\cite{Mustapha2012} and~\cite{Manard2017all}, crosses -- the scalar-relativistic theoretical calculations from~\cite{JCP52,IJMS2}. Results that include SO coupling~\citep{IJMS2} are shown by pluses. Asterisks present the ASR theoretical results for $_{71}^{175}$Lu$^+$ (4f$^{14}$5d6s, $^3$D). The labels mark electronic configurations of ions that differ from the most common 4f$^m$6s. }
\label{fig:Overview}
\end{figure}

Figure~\ref{fig:Overview} provides an overview of the room-temperature zero-field mobilities available for lanthanide ions. Most of the ions have the 4f$^m$6s ground-state configuration and their mobilities follow well-defined trend lines. Remarkable deviations take place for a few ions with different outer shell occupancies: Ce$^+$ (5d$^2$), Gd$^+$ (5d6s) and Lu$^+$ (6s$^2$). Noteworthy, theory predicts quite similar mobilities for Gd$^+$ and Lu$^+$ in the metastable $^3$D state of the same 5d6s configuration, whereas the difference in the ground- and metastable-state Lu$^+$ mobilities is huge comparing to the trend line variation. This clearly confirms the sensitivity of the ion transport to the ion electronic configuration that underlies the electronic state chromatography effect~\citep{Kemper1991, Bowers1993, Taylor1999, Iceman2007, Ibrahim2008, Manard2016b, Manard2016a}. On the other hand, the striking difference between the mobilities in He and Ar gases looks surprising, not because of the magnitudes of the $K_0$ values (which arise due to the ion-neutral reduced masses and interaction strengths), but because the trends with atomic number are so different. While the mobility of the 4f$^m$6s ions in He generally increases with $Z$, that in Ar remains almost constant. Moreover, the change of electronic configuration causes opposite mobility variations in the two gases. This behavior can only be understood by analyzing the features of the ion-atom interactions and their manifestations in the transport properties. To justify such an analysis, we should emphasize the very good agreement between the experimental and theoretical data shown in figure~\ref{fig:Overview}. The most remarkable exception of the $_{64}^{156}$Gd$^+$ (4f$^7$5d6s) ion originates in fact from the vectorial SO coupling effect~\citep{IJMS2}. Compared to the small-core CCSD(T) results, the potentials obtained by the MRCI method lack the accuracy required for transport calculations~\citep{JCP52}. Consideration of $_{63}^{151}$Eu$^+$ (4f$^7$6s) revealed worse performance of the large-core description of lanthanide ions within the CCSD(T) framework~\citep{JCP52,TimEd}.

\subsection{Interaction potentials}
\label{ssec:Lapots}

Here, we provide a brief presentation of the {\it ab initio} approach that was successfully applied for the lanthanide ions to help understanding its extension to the actinide ions, where no direct comparison with experiment is possible so far (see below). It relies on the small-core (28 electron) ECPs adjusted at the quasi-relativistic, Wood-Boring, Hartree-Fock level of theory, ECP28MWB~\citep{ECP28MWB}. The supplementary atomic natural orbital basis sets~\citep{LaANO} suffer from the lack of diffuse functions. The optimized s2pdfg diffuse augmentation~\citep{SC1} was therefore used together with the segmented basis contraction~\citep{LaSegm}. He and Ar atoms were described using the augmented, correlation-consistent, polarized basis sets aug-cc-pV5Z~\citep{AVQZHe} and the 3s3p2d2f1g bond function set~\citep{Bf} was placed at the midpoint of the ion-atom distance. The CCSD(T) calculations were performed using the Hartree-Fock references and kept the 4s$^2$4p$^6$4d$^{10}$ shells of the ion and the 1s$^2$2s$^2$2p$^6$ shells of the Ar atom in core. \hl{For the states of D symmetry (Gd$^+$ and Lu$^+$ ions)} Hartree-Fock reference wave functions were obtained for each $\Lambda$ separately, using either different symmetry representations or enforcing an electron population of the  5d$_\sigma$ or 5d$_\delta$ orbital as described by~\cite{IJMS2}. The CCSD(T) potentials were obtained on fine grids of internuclear distances extending up to 25-40 {\AA} and corrected for basis set superposition error by means of the counterpoise procedure by~\cite{BoysB}. The MOLPRO program package~\citep{MOLPRO} was used for all calculations.

\begin{figure}[hbt!]
\includegraphics[width=1.0\linewidth]{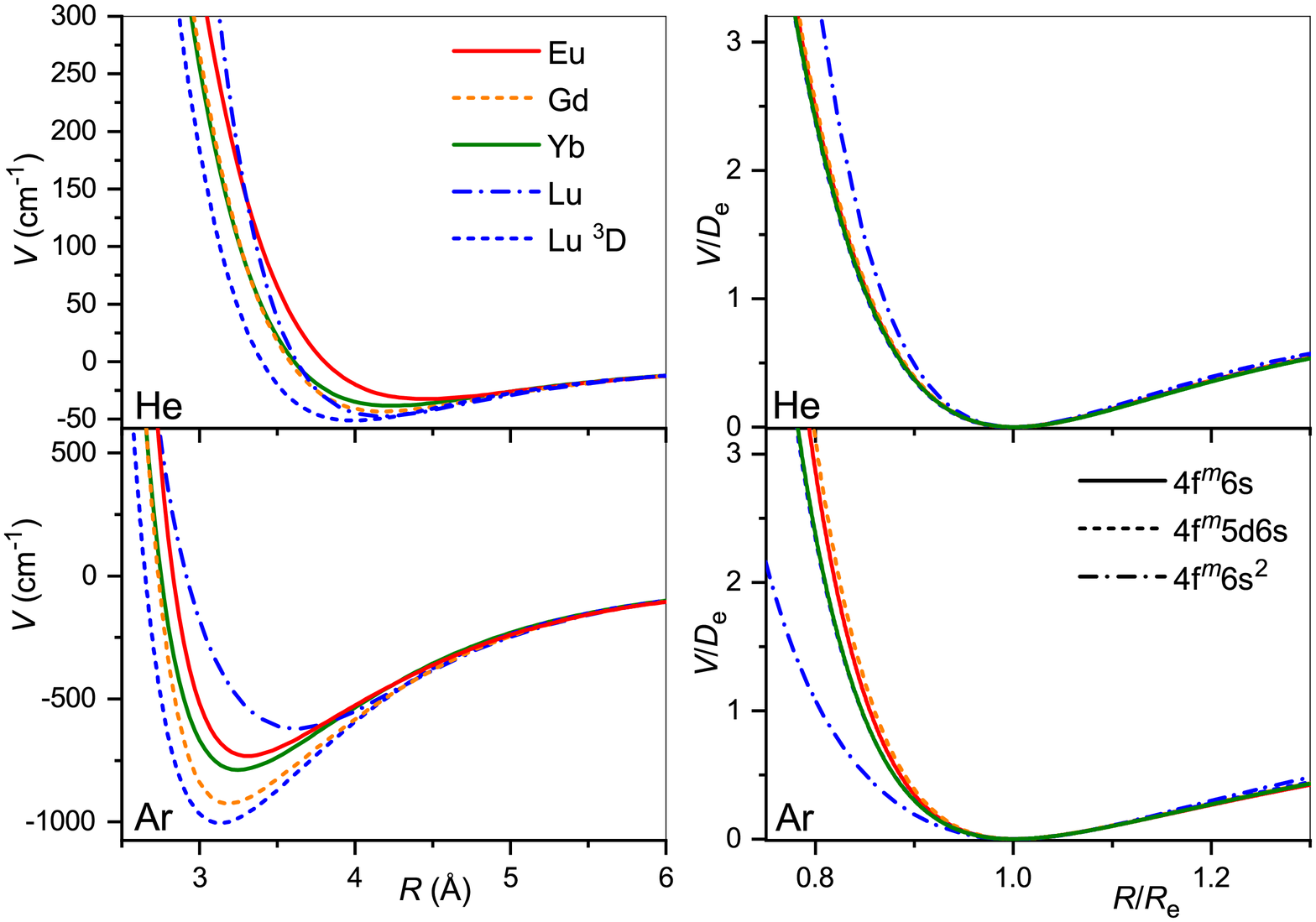}
\caption{Interaction potentials of the lanthanide ions with He (top panels) and Ar (bottom panels). True and reduced potentials are shown on the left and on the right, respectively. }
\label{fig:Lapots}
\end{figure}

\begin{table}[hbt!]
\caption{Equilibrium parameters of the ion-atom interaction potentials for lanthanide ions, $R_e$ ({\AA}) and $D_e$ (cm$^{-1}$).} \label{tab:equil}
  \begin{center}
 \begin{threeparttable}
  \begin{tabular}{l@{\quad}c@{\quad}c@{\qquad}c@{\quad}c@{\quad}}
\hline
\hline
Ion & \multicolumn{2}{c}{He} & \multicolumn{2}{c}{Ar} \\
\cline{2-5}
      & $R_e$ & $D_e$ & $R_e$ & $D_e$ \\
\hline
\hline
Eu$^+$ 4f$^7$6s $^9$S$^\circ$ & 4.45 & 33 & 3.31 & 732 \\
Gd$^+$ 4f$^7$5d6s $^{10}$D$^\circ$~\tnote{a} & 4.18 & 43 & 3.18 & 925 \\
Yb$^+$ 4f$^{14}$6s $^2$S & 4.23 & 38 & 3.25 & 789 \\
Lu$^+$ 4f$^{14}$6s$^2$ $^1$S & 4.17 & 47 & 3.62 & 620 \\
Lu$^+$ 4f$^{14}$5d6s $^3$D~\tnote{a} & 3.99 & 51 & 3.13 & 1005 \\
\hline
\hline
  \end{tabular}
     \begin{tablenotes}
\item[a] Parameters of the isotropic potential $V_0$
   \end{tablenotes}
  \end{threeparttable}
  \end{center}
  \end{table}

The  obtained interaction potentials are plotted in the left column of figure~\ref{fig:Lapots}, while \hl{the parameters of their minima, equilibrium distances $R_e$ and binding energies $D_e$, are presented} in table~~\ref{tab:equil}. Tabulated potential functions are given in the LXCat database~\citep{LXCAT}. Note that \hl{for the  Gd$^+$ ($^{10}$D$^\circ$) and Lu$^+$ ($^3$D),} only the isotropic potentials $V_0(R)$, eq.~(\ref{eq:v0}), are discussed hereafter. The lowest-order induction interaction, $V_\mathrm{ind}(R)=-\alpha_\mathrm{RG}/2R^4$, where $\alpha_\mathrm{RG}$ is the static dipole polarizability of \hl{the rare gas atom (RG)}, does not depend on the nature of the ion and determines the common features of interaction potentials at large separations. Indeed, the deviations of $D_e$ from $V_\mathrm{ind}(R_e)$ do not exceed 20\%, being generally positive (more attraction) for He and negative (more repulsion) for Ar. \hl{Equilibrium distance generally decreases with $Z$ for ions with the same valence electronic configurations.  In contrast, population of the 5d shell enhances the interaction energy and shrinks the equilibrium distance.}

It is instructive to compare the overall shapes of the potentials by introducing the reduced functions, $V(R/R_e)/D_e$, as depicted in the right column of figure~\ref{fig:Lapots}. In the case of He, \hl{the reduced potentials are hardly distinguishable from each other} except that for the Lu$^+$ ion with its unique closed-shell, 6s$^2$ configuration. The reduced potentials show an exception for Lu$^+$  with Ar too, but now with a softer repulsive wall. \hl{In contrast to He case, repulsive interaction of the Eu$^+$, Gd$^+$ and Yb$^+$, Lu$^+$($^3$D) ions with Ar differ slightly from each other. This reflects the effect of the 4f$^7$ and 4f$^{14}$ occupancies.}

\subsection{Ion mobility}
\label{ssec:Lamobs}

\begin{figure}[hbt!]
\includegraphics[width=1.0\linewidth]{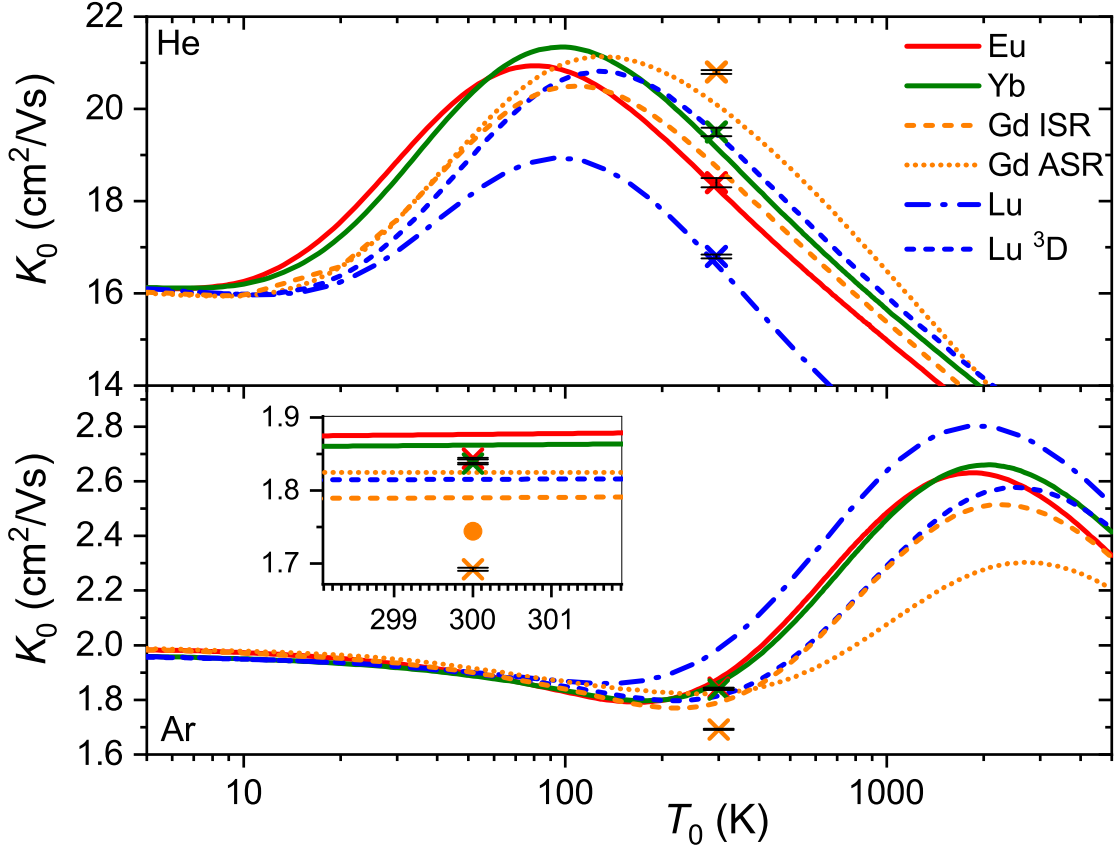}
\caption{Zero-field mobilities of the lanthanide ions in He (top panel) and Ar (bottom panel) calculated as functions of temperature. Crosses \hl{with tiny error bars} indicate experimental data by~\cite{Mustapha2012} and~\cite{Manard2017all}. The inset for Ar provides an enlarged view of the room-temperature region. The dot represents an accurate calculation~\citep{IJMS2} for the $_{64}^{156}$Gd$^+$ (4f$^7$5d6s, $^{10}$D$^\circ_{5/2}$) ion that includes the vectorial SO coupling and hence emphasizes the error of the scalar relativistic approach. The same example is used to illustrate the difference between ISR and ASR approximations.}
\label{fig:Lamob}
\end{figure}

While the good agreement between the experimental and theoretical mobilities at room temperature demonstrated in figure~\ref{fig:Overview}  indicates reasonable accuracy of the scalar-relativistic {\it ab initio} interaction potentials, only a wide temperature dependence of the mobility can fully uncover the features pertinent to a particular ion-neutral interaction~\citep{MasonBook1988,LarryBook2018}. Unfortunately, no such measurements have been performed so far for the lanthanide ions, so only theoretical dependences are available (see figure~\ref{fig:Lamob}).

Our first comment is that the ISR approximation does not work well for the $_{64}^{156}$Gd$^+$ (4f$^7$5d6s) ion~\citep{IJMS2}. In other words, its transport properties are not reproduced quantitatively by a single isotropic $V_0$ potential. By contrast, the ISR and ASR (not shown) approximations agree with each other well for the $_{71}^{175}$Lu$^+$ (4f$^{14}$5d6s) ion. The second comment is that the main features of the interaction potentials discussed above are clearly reflected in $K_0(T_0)$. Those for $_{63}^{151}$Eu$^+$ (4f$^7$6s) and $_{70}^{174}$Yb$^+$ (4f$^{14}$6s), and for $_{64}^{156}$Gd$^+$ (4f$^7$5d6s) (ISR approximation) and $_{71}^{175}$Lu$^+$ (4f$^{14}$5d6s), pair with each other, whereas the results for $_{71}^{175}$Lu$^+$ (4f$^{14}$6s$^2$) differ, like the underlying potentials do. Third, it is evident that room temperature corresponds to different regions of the mobility functions for He and Ar. In the former case, it falls beyond the mobility maximum, while in the latter it falls in the region of the shallow mobility minimum. A rich early-days experience with model potential functions and direct potential inversion~\citep{Viehland1976,Viehland1983}, summarized by~ \citet*[chap.7]{MasonBook1988} and \citet*[chap.9]{LarryBook2018}, helps to connect, qualitatively, the radial dependence of the potential and temperature dependence of the mobility. The low-temperature trend towards the polarization limit reflects the dominant interaction term, $V_\mathrm{ind}$, the mobility minimum features an intermediate interaction range where attractive van der Waals forces of higher order are also operative, the maximum is predominantly connected to the potential well, and the decreasing high-temperature branch reflects the repulsive interactions. Note that flipping a $K_0(T_0)$  plot upside down and right to left, one sees a cartoon of an interaction potential. An immediate conclusion is that the room temperature measurements in different buffer gases do not equally attest the properties of the ion. These and the ion-neutral reduced masses provide a good but partial explanation for why the variations of the room-temperature mobility with the nature of ion are so strikingly different for He and Ar (see figure~\ref{fig:Lapots}). However, the electronic state difference also contributes substantially (figure~\ref{fig:Lamob}). As a side note, the slightly deeper mobility minimum in Ar can be mentioned for the reduced potentials of Gd$^+$ and Lu$^+$ (4f$^{14}$5d6s); these may reflect an interaction of the ion permanent quadrupole moment with the induced dipole moment of an atom, which is obviously absent for the S-state ions~\citep{Breckenridge2002}.

\subsection{Sensitivity to electronic configuration}
\label{ssec:LaESC}

One way to quantify the mobility variations with the electronic configuration of the ion can be closely related to so-called electronic state chromatography effect, or the discrimination of the ground- and metastable-state ions by distinct mean drift times. Although well studied experimentally for the transition metal ions~\citep{Kemper1991, Bowers1993, Taylor1999, Iceman2007, Ibrahim2008, Manard2016b, Manard2016a}, this effect has not been investigated for other ions. Theoretical results for Lu$^+$ (4f$^{14}$5d6s) allow us to demonstrate this effect for the lanthanide family.

\begin{figure}[hbt!]
\centering
\includegraphics[width=0.7\linewidth]{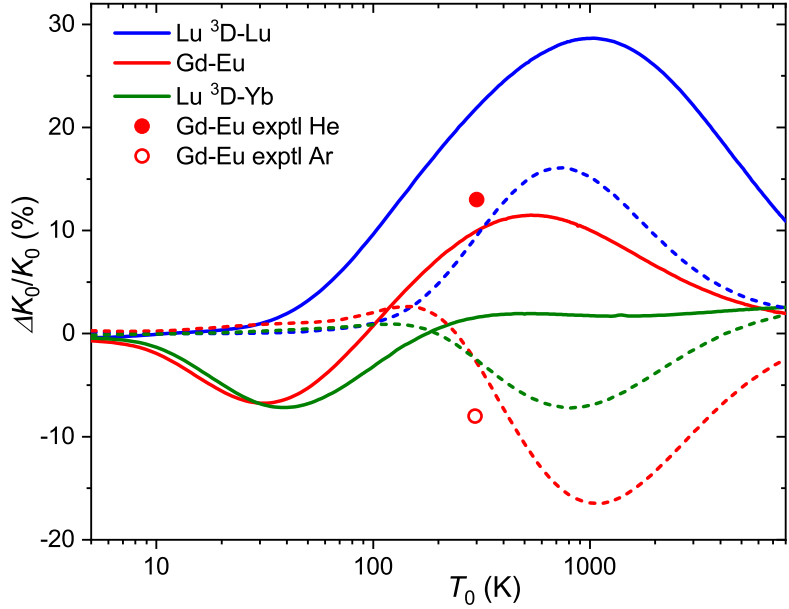}
\caption{Relative changes in the ion mobilities upon 6s$^2 \to$ 5d6s excitation of the Lu$^+$ ion and upon ``adding'' d electron to Eu$^+$ and Yb$^+$ ions.  Solid and dashed lines are used for He and Ar buffer gases, respectively. Experimental room-temperature values are derived from~\cite{Mustapha2012} and~\cite{Manard2017}.  }
\label{fig:DKK}
\end{figure}

It is convenient to consider the drift time of the ion given by eq. (\ref{eq:td}). Marking the quantities related to metastable ions by an asterisk and using eq.  (\ref{eq:K0}), one gets
\begin{equation}\label{eq:deltaTabs}
\Delta t_d^* = t_d^* - t_d =
\frac{1}{E/n_0}\frac{l}{N_L}\frac{K_0-K^*_0}{K^*_0K_0}
\end{equation}
for the absolute drift time difference and
\begin{equation}\label{eq:deltaTrel}
\Delta t_d^*/t_d^* = (K_0-K^*_0)/K_0 = -\Delta K_0/K_0
\end{equation}
for the relative one, where $\Delta K_0/K_0$ is the relative deviation of the mobility of the metastable state ion from that of the ground-state ion. Note that it depends on temperature (and $E/n_0$) through the individual mobilities. Figure~\ref{fig:DKK} shows the zero-field $\Delta K_0/K_0$ ratios for $_{71}^{175}$Lu$^+$ (4f$^{14}$6s$^2$, $^1$S) and $_{71}^{175}$Lu$^+$ (4f$^{14}$5d6s, $^3$D) ions as a function of temperature. The maximum difference in drift times in He and Ar amounts 30 and 15\% at 750 and 1000 K, respectively. The room-temperature difference in He, 22\%, is comparable to those measured~\citep{Ibrahim2008} for the coinage metal ions (50, 25 and 13\% for copper, silver and gold). These values attest the discrimination between the ground $n$d$^{10}$ and metastable $n$d$^9$($n+1$)s configurations. Comparison for the third-row transition metal ions that have 6s and 6s$^2$ ground and metastable configurations, like Hf$^+$, Re$^+$ or Hg$^+$, would be more relevant to Lu$^+$ ion, but to our knowledge none of these ions has been detected in metastable states  in mobility experiments~\citep{Taylor1999}.

The same pictorial approach can be used for the mobilities of distinct ions in similar configurations. From the present data, the effect of ``adding'' a 5d electron to the 6s one can be viewed for the Gd$^+$-Eu$^+$ and Lu$^+$($^3$D)-Yb$^+$ pairs. The corresponding $\Delta K_0/K_0$ ratios (5d6s ion is taken \hl{as the ``metastable'' state}) are also plotted in figure~\ref{fig:DKK}. In general, they follow a similar trend for each buffer gas, but the two trends are almost opposite. Interestingly, the calculated mobilities demonstrate that higher sensitivity to electronic configuration can sometimes be achieved in Ar rather than He.

\subsection{Ionic radii}
\label{ssec:Larad}

Effective ionic radii are important parameters in crystallography, electronic structure theory and molecular modeling. For heavy ions, their dependence on $Z$ should reveal the effect of relativistic contraction. Though the effective size of an ion can be extracted from the {\it ab initio} interaction potentials themselves, it is important to understand whether or not the transport
measurements can provide a systematic means to probe the ionic radii, taking into account exploratory experiments for the actinide ions~\citep{Sewtz2003,Backe2005} and speculating to the superheavy ions.

The ionic radius can be defined simply as
\begin{equation}
\label{eq:radWB}
R_\mathrm{ion}(R) = R_e - R_\mathrm{RG},
\end{equation}
where $R_e$ is the equilibrium distance of the ion-RG interaction potential and $R_\mathrm{RG}$ is the atomic radius of the RG atom, here He or Ar. This definition was analyzed by~\cite{WrightBreckenridge2010}  (WB), who recommended the systematics based on He interactions (with the van der Waals radius of 1.49 {\AA}) and noticed that significant distortions of an ion electron density by Ar ($R_\mathrm{Ar}=1.88$ {\AA}) make the definition (\ref{eq:radWB}) inconsistent for that RG.

To deduce $R_\mathrm{ion}$  (or, equivalently, $R_e$) from the zero-field mobility, one should use eq. (\ref{eq:Langevin}) and somehow relate $\bar{\Omega}^{(1,1)}(T_0)$ to the ion-neutral interaction potential. Within the hard sphere (HS) model
\begin{equation}
\label{eq:HSOmega}
\bar{\Omega}^{(1,1)}(T_0) = \pi R_e^2.
\end{equation}
Combining eqs. (\ref{eq:Langevin}) and (\ref{eq:HSOmega}), one finds that
\begin{equation}
\label{eq:HSRe}
R_e = \left(\frac{2}{\pi\mu_0k_BT_0}\right)^{1/4}\left(\frac{3q}{16N_0K_0(T_0)}\right)^{1/2}.
\end{equation}
Then one can easily obtain $R_\mathrm{ion}$ from eq. (\ref{eq:radWB}).

\begin{figure}[hbt!]
\includegraphics[width=1.0\linewidth]{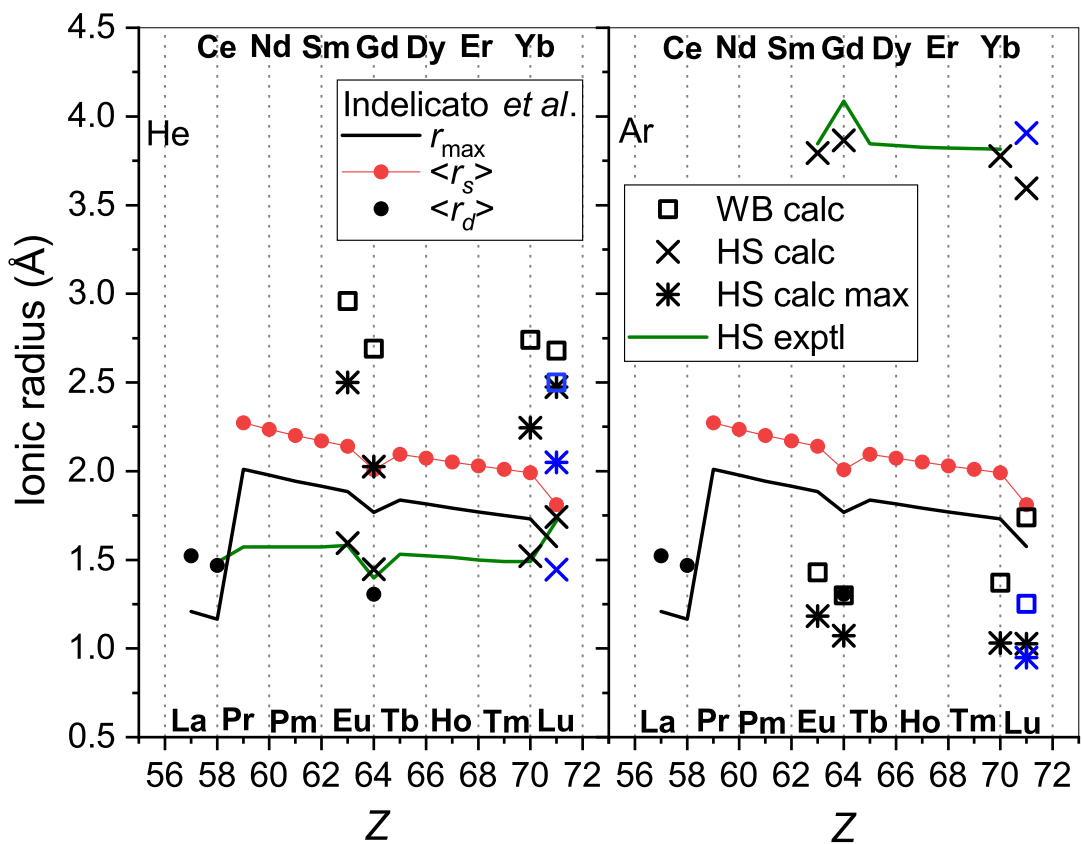}
\caption{
The radii of the lanthanide ions determined from He (left panel) and Ar (right panel) data. Presented are the WB radii from {\it ab initio} calculations and the results of the HS model applied to experimental and calculated room-temperature mobilities \hl{``HS exptl'' and ``HS calc'', respectively)} and to the calculated mobility at its maximum \hl{(``HS calc max'')}. Blue color is used for $_{71}^{175}$Lu$^+$ (4f$^{14}$5d6s). Parameters of the ion electron distributions calculated by~\cite{Indelicato2007} are also shown; see text for explanation.}
\label{fig:Larad}
\end{figure}

In figure~\ref{fig:Larad} radii obtained this way are compared with the parameters of the radial electron distributions calculated by~\cite{Indelicato2007} for bare ions, namely, the mean radii of the 6s orbitals, $<r_s>$, and (if any) the 5d orbitals, $<r_d>$, and with the maximum of the density of the outermost orbital, $r_\mathrm{max}$. The case of He provides a quite consistent picture. The WB radii correlate well (within 15\%) with the electronic parameters, being 1 and 0.8 {\AA} larger than $r_\mathrm{max}$  and $<r_s>$, respectively, due to the presence of the He atom.

Effective contraction of the bare ion radius in He when going from Eu$^+$ to Yb$^+$ amounts to 0.15 {\AA}, whereas the WB radius shrinks by 0.22 {\AA}. \hl{In contrast, the HS model applied to both experimental and calculated  room-temperature mobility data gives smaller radii and underestimates their contraction} (0.09 {\AA} for the Yb$^+$-Eu$^+$ pair). When applied to the theoretical mobilities at their maxima, \hl{the HS model gives a more consistent trend; results become closer to the WB definition and the Yb$^+$-Eu$^+$ contraction becomes 0.26 {\AA}. Still, the HS model works reasonably only for potentials of very similar shape. Even a minor deviation at the repulsive wall in the case of Lu$^+$-He interaction} (see figure~\ref{fig:Lapots}) causes an artificial increase of the effective radius.

In the case of Ar as the buffer gas, there is a much larger mismatch between the electronic parameters and models based on ion-atom interactions and transport. The effective ionic radii derived from interaction potentials are too small in comparison to $<r_s>$ and even $r_\mathrm{max}$, show weaker $Z$-dependence and opposite variation for the ``soft'' Lu$^+$-Ar interaction. This is in line with the analysis by~\cite{WrightBreckenridge2010} for lighter ions. The HS model works reasonably for mobilities at their maxima but gives meaningless results for room-temperature mobilities.

\section{Actinide ions}
\label{sec:ac}

The data on actinide ion mobility are very scarce. In fact, the only dedicated experiment is that of~\cite{Biondi1972} who measured the mobility of $_{92}^{238}$U$^+$ (5f$^3$7s$^2$) in He as a function of $E/n_0$ and pressure near room temperature. Smoothed data are tabulated by~\cite{EllisADNDT} and are also available in the LXCat database~\citep{LXCAT}. A few relative drift time measurements were carried out~\citep{Sewtz2003,Backe2005} in Ar for the $_{100}^{255}$Fm$^+$ (5f$^{12}$7s):$_{98}^{251}$Cf$^+$ (5f$^{10}$7s) and $_{95}^{243}$Am$^+$ (5f$^7$7s):$_{94}^{239}$Pu$^+$ (5f$^6$7s) ion pairs to assess their effective radii. Mobilities of U$^+$ ion in all rare gases from He to Xe were also calculated using {\it ab initio} interaction potentials~\citep{TimUplus}.

\subsection{Interaction potentials}
\label{ssec:acpots}

\begin{table}[hbt!]
\caption{Equilibrium parameters of the ion-atom interaction potentials for actinide ions, $R_e$ ({\AA}) and $D_e$ (cm$^{-1}$).} \label{tab:Acpot}
  \begin{center}
 \begin{threeparttable}
  \begin{tabular}{l@{\quad}c@{\quad}c@{\qquad}c@{\quad}c@{\quad}}
\hline
\hline
Ion & \multicolumn{2}{c}{He} & \multicolumn{2}{c}{Ar} \\
\cline{2-5}
      & $R_e$ & $D_e$ & $R_e$ & $D_e$ \\
\hline
\hline
Ac$^+$ 7s$^2$ $^1$S & 4.82 & 30 & 4.07 & 426 \\
Ac$^+$ 7s$^2$ $^1$S~\tnote{a} & 4.80 & 30 & 4.04 & 434 \\
U$^+$ [5f$^3$]7s$^2$~\tnote{b} & 4.62 & 33 & 3.96 & 454 \\
U$^+$ [5f$^3$]7s$^2$~\tnote{a} & 4.59 & 34 & 3.96 & 470 \\
Am$^+$ 5f$^7$7s $^9$S$^\circ$ & 4.27 & 39 & 3.45 & 698 \\
Cm$^+$ 5f$^7$7s$^2$ $^8$S$^\circ$ & 4.36 & 42 & 3.82 & 538 \\
Cm$^+$ [5f$^7$]7s$^2$~\tnote{a} & 4.39 & 40 & 3.88 & 509 \\
No$^+$ 5f$^{14}$7s $^2$S & 4.03 & 48 & 3.38 & 763 \\
Lr$^+$ 5f$^{14}$7s$^2$ $^1$S & 4.08 & 52 & 3.71 & 598 \\
Lr$^+$ [5f$^{14}$]7s$^2$~\tnote{a} & 4.11 & 50 & 3.78 & 565 \\
\hline
\hline
  \end{tabular}
     \begin{tablenotes}
\item[a] Large-core calculations, this work.
\item[b] Large-core calculations by~\cite{TimUplus}.
   \end{tablenotes}
  \end{threeparttable}
  \end{center}
  \end{table}

\hl{Accepting scalar relativistic approximation for actinide ions, one can straightforwardly extend the} {\it ab initio} approach described above for the lanthanide family. Instead of the small-core 28-electron ECP28MWB effective core potentials, compatible 60-electron ECP60MWB ones~\citep{Kuchle1994} with analogous segmented atomic natural orbital basis sets~\citep{AcSegm}  have to be used. The exponents of the spdfg set of the diffuse primitives (0.01, 0.008, 0.03, 0.07 and 0.05, respectively) have been optimized for the polarizabilities of the neutral Am, No and Lr atoms. The basis sets for He and Ar and other features of the CCSD(T) calculations remain the same. The single-reference restriction limits the application of this approach to the ground states of the $_{89}^{227}$Ac$^+$ (7s$^2$, $^1$S), $_{95}^{241}$Am$^+$ (5f$^7$7s, $^9$S$^\circ$), $_{96}^{247}$Cm$^+$ (5f$^7$7s$^2$, $^8$S$^\circ$), $_{102}^{254}$No$^+$ (5f$^{14}$7s, $^2$S) and $_{103}^{255}$Lr$^+$ (5f$^{14}$7s$^2$, $^1$S) ions.

\begin{figure}[hbt!]
\includegraphics[width=1.0\linewidth]{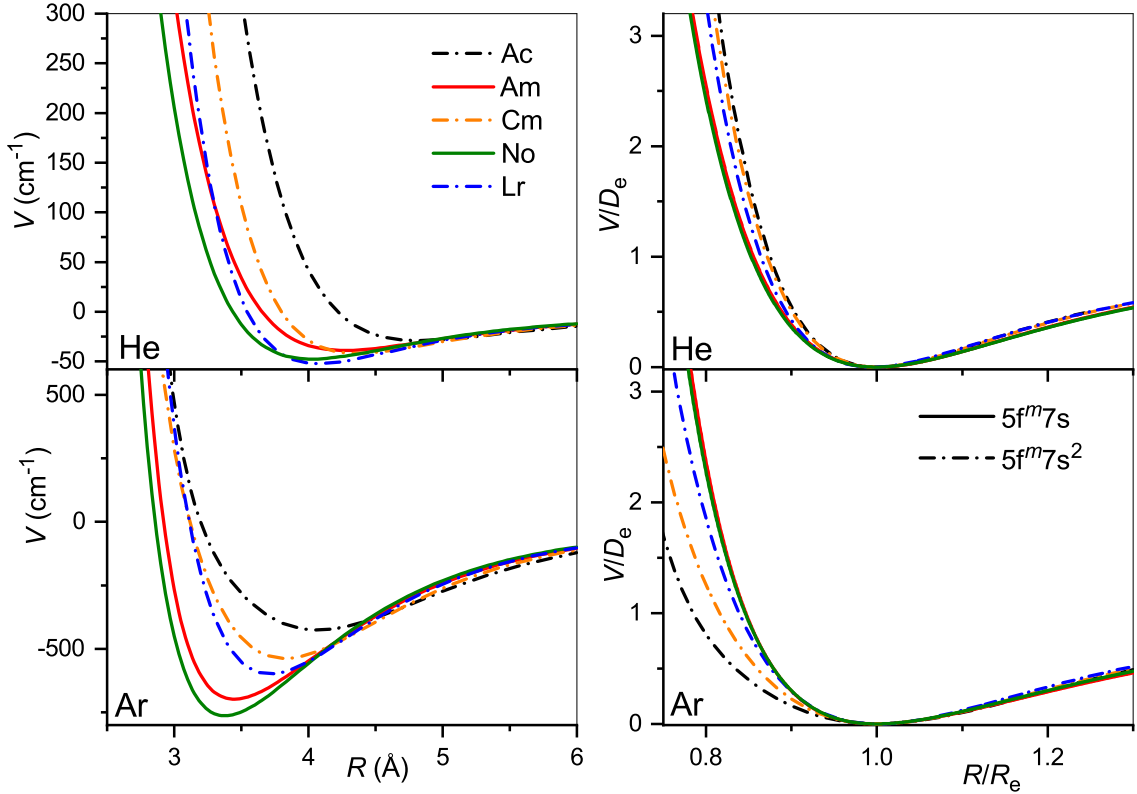}
\caption{Interaction potentials of the actinide ions with He (top panels) and Ar (bottom panels). True and reduced potentials are shown on the left and on the right, respectively.}
\label{fig:Acpots}
\end{figure}

\begin{figure}[hbt!]
\includegraphics[width=1.0\linewidth]{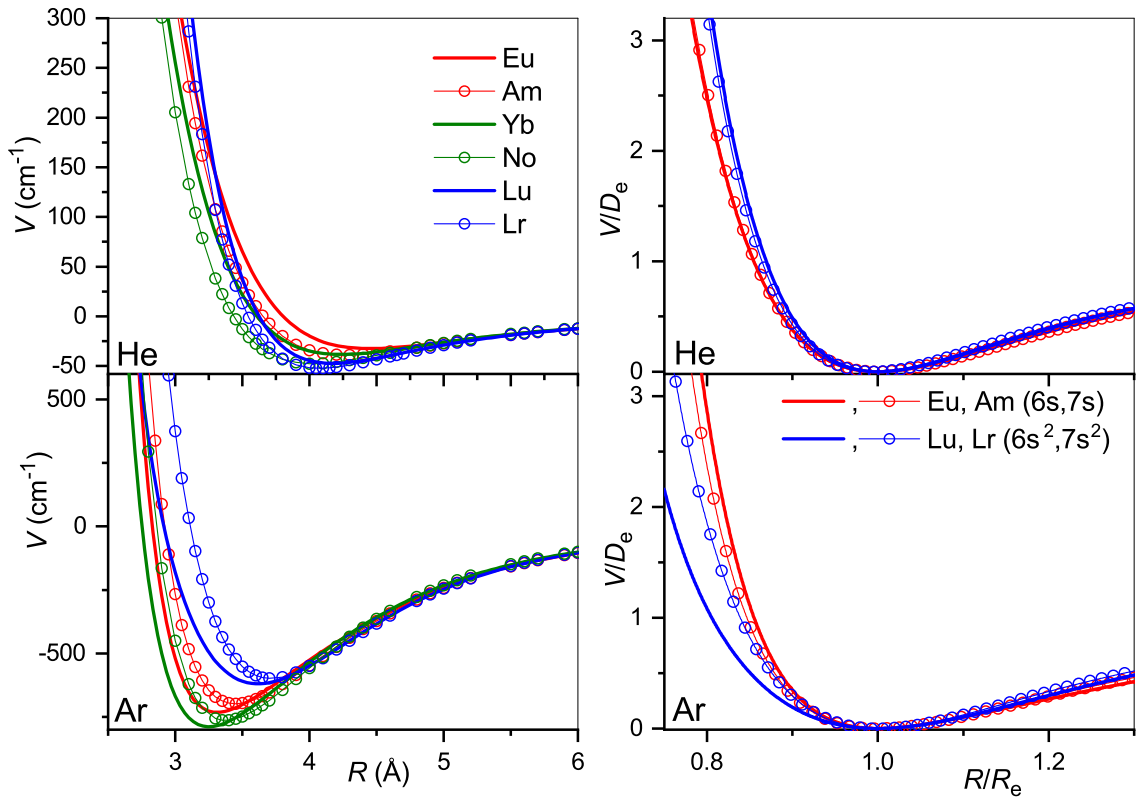}
\caption{Interaction potentials of the analogous actinide and lanthanide ions with He (top panels) and Ar (bottom panels). True and reduced potentials are shown on the left and on the right, respectively.  }
\label{fig:AcLapots}
\end{figure}

An alternative approach was suggested by~\cite{TimUplus}. It exploits large-core ECPs (``5f-in-core'') that absorb 5s5p5d5f shells leaving for explicit consideration the outer 6s, 6p, 6d, 7s,... electrons only~\citep{LCECP}. With this approach, electronic angular momenta and configuration mixing effects due to incomplete 5f shell occupancies are hidden in the ECPs and the ground electronic states of the ions acquire $^1$S or $^2$S effective symmetry, except for $^{232}$Th$^+$ (6d$^2$7s). To test the difference with the small-core approach in a systematic way, we calculated the CCSD(T) interaction potentials for the ground-state Ac$^+$, U$^+$, Cm$^+$ and Lr$^+$ ions with He and Ar using the same large-core ECPs. In contrast to~\cite{TimUplus}, we used the supplementary basis sets of aug-cc-pVQZ quality without further modification but augmented by the 3s3p2d2f1g \hl{bond function} set~\citep{Bf} placed in the middle of the ion-atom distance. The results are shown in table~\ref{tab:Acpot}.

For the Ac$^+$ ion without 5f electrons, the comparison apparently favors the large-core description that gives slightly stronger ion-atom interactions. However, the opposite is seen for the 5f$^7$ and 5f$^{14}$ configurations of Cm$^+$ and Lr$^+$ ions. The interaction strengths differ by 4-5\% for He and by 5-6\% in Ar, whereas the equilibrium distances differ by 0.03-0.06 {\AA}. A reason for caution with the large-core approach is its modest accuracy for mobility calculations of U$^+$ in He~\citep{TimUplus} and for Eu$^+$ in He~\citep{JCP52} (with analogous ``4f-in-core'' ECP); these were low by 8 and 4\%, respectively, compared to the experimental values.  Also, for open-shell ions it permits only the simplest ISR calculation for collision cross sections and transport properties. In what follows we will consider only the small-core approach, since it is consistent with the lanthanide results summarized above.

The true and reduced interaction potentials for the actinide ions are shown in figure~\ref{fig:Acpots}. As in the lanthanide case shown in figure~\ref{fig:Lapots}, interactions of the actinide ions with 7s and 7s$^2$ outer shells differ significantly from each other. They exhibit weaker bonding and repulsion that is stronger for He and softer for Ar. The dependence on the inner f-shell occupancy is more pronounced than in lanthanides, in accord with the facts known from chemical interactions. Actinide ions with the 7s configuration interact with He more strongly than their lanthanide counterparts, with $R_e$ reduced by almost 0.2 {\AA} and $D_e$ increased by more than 20\%. In contrast, $R_e$ increases when switching from Lu$^+$ to Lr$^+$ ion with the $n$s$^2$ configuration being accompanied by a marginal 2\% increase of the binding energy. Interactions with Ar are weaker for actinide ions regardless of the outer configuration. Overall, the two ion families demonstrate impressive similarity in their interaction potentials. This is illustrated in figure~\ref{fig:AcLapots} that presents the potentials for various analogs. Especially telling are the reduced potentials showing that the difference due to outer $n$s occupancy decreases from the lanthanides to the actinides. Note that reduced potentials for the No$^+$ and Yb$^+$ are indistinguishable from those of Am$^+$ and Eu$^+$ at the scale of the figure.

\subsection{Ion mobility}
\label{ssec:Acmobs}

\begin{figure}[hbt!]
\includegraphics[width=1.0\linewidth]{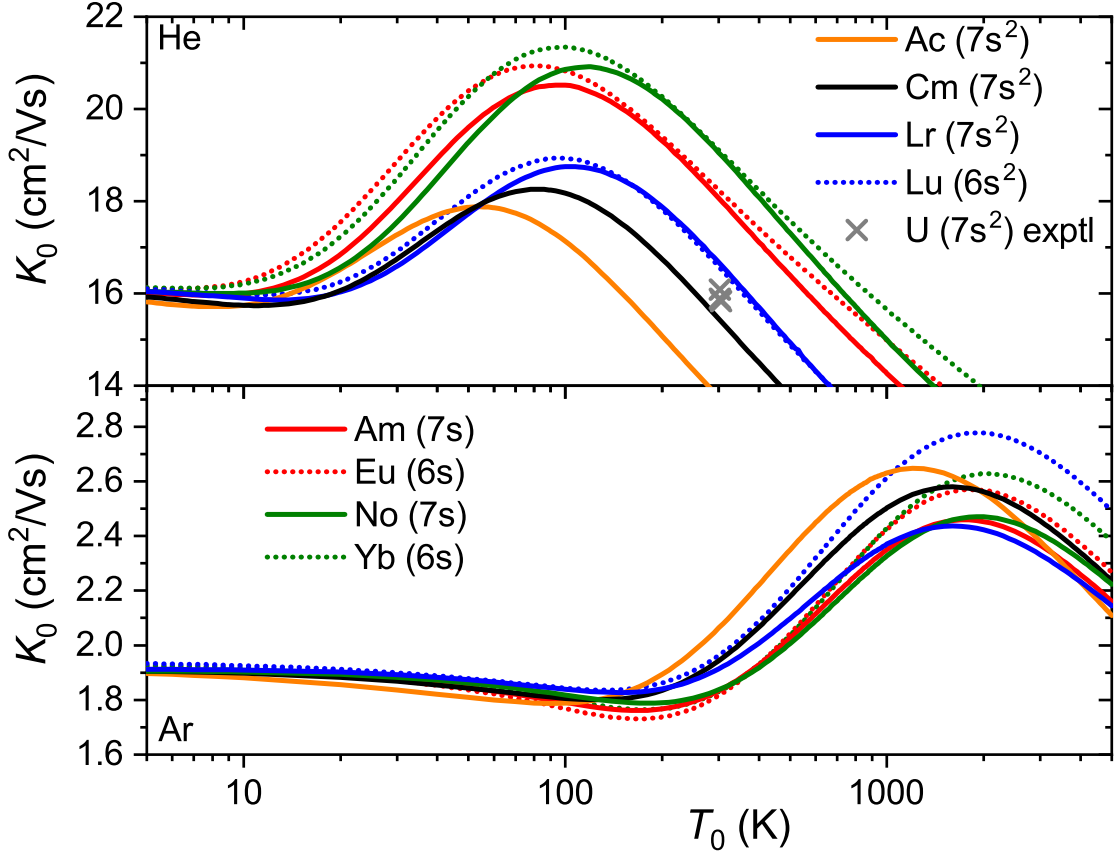}
\caption{Zero-field mobilities of some actinide ions in He (top panel) and Ar (bottom panel) calculated as functions of temperature. For comparison, mobilities of the lanthanide analogs are also shown. Crosses indicate experimental data~\citep{Biondi1972} for the U$^+$ ion in He.}
\label{fig:Acmob}
\end{figure}

The interaction potentials described above were used to compute the mobilities of $_{89}^{227}$Ac$^+$ (7s$^2$), $_{95}^{241}$Am$^+$ (5f$^7$7s), $_{96}^{247}$Cm$^+$ (5f$^7$7s$^2$), $_{102}^{254}$No$^+$ (5f$^{14}$7s) and $_{103}^{255}$Lr$^+$ (5f$^{14}$7s$^2$) in He and Ar. The calculated temperature dependences shown in figure~\ref{fig:Acmob} exhibit trends similar to those found in lanthanides. The mobility maxima in He for ions with both 7s and 7s$^2$ configurations are slightly reduced and shifted towards higher temperatures. The trend of increasing mobility with $Z$ is visible for ions of both groups, Am$^+$-No$^+$ and Ac$^+$-Cm$^+$-Lr$^+$. Experimental data by~\cite{Biondi1972} for U$^+$ in He, though somewhat uncertain, does not support the latter trend, but fits to the theoretical results for 7s$^2$ group. In Ar, the mobility of the ions with 7s$^2$ configuration follows the reverse trend, decreasing along the Ac$^+$-Cm$^+$-Lr$^+$ sequence, while the difference in Am$^+$ and No$^+$ mobilities becomes marginal. As discussed above, such a reversal also takes place for lanthanide ions of the 5d6s configuration and, similarly, mirrors the short-range behavior of ion-atom interaction potentials.

\subsection{Sensitivity to electronic configuration}
\label{ssec:AcESC}

\begin{figure}[hbt!]
\centering
\includegraphics[width=0.7\linewidth]{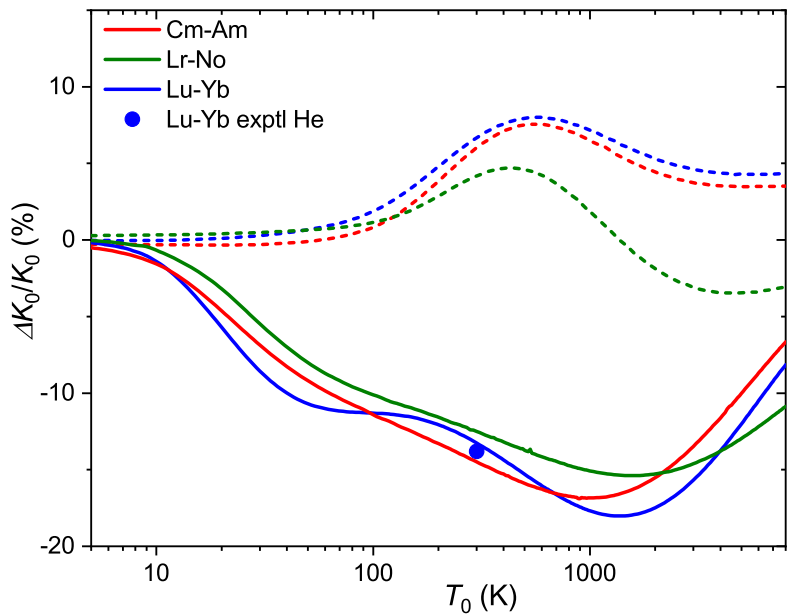}
\caption{Relative changes in the ion mobilities between 7s$^2$ and 7s ions Cm$^+$-Am$^+$ and Lr$^+$-No$^+$. The lanthanide analog of the latter pair,  Lu$^+$-Yb$^+$, is also shown. Solid and dashed lines are used for He and Ar buffer gases, respectively. The experimental room-temperature value is from~\cite{Manard2017}. }
\label{fig:DKKAc}
\end{figure}

In  figure~\ref{fig:DKKAc} are plotted the relative mobility differences, $\Delta K_0/K_0$, for the Cm$^+$-Am$^+$ and Lr$^+$-No$^+$ pairs of ions that differ by their 7s occupancies in comparison with that for lanthanide analog Lu$^+$-Yb$^+$. In He, all three pairs behave similarly, giving room-temperature drift time difference of 10-15\%. As has been already mentioned, the difference in the mobility of 7s and 7s$^2$ ions in Ar has the opposite sign. Interestingly, the difference due to 5f shell occupancy between Cm$^+$-Am$^+$ and Lr$^+$-No$^+$ is larger than that between the lanthanide and actinide families.

Overall, the effect on the mobility in both buffer gases of outer $n$s shell occupancy in the lanthanide and actinide ions is smaller than the effect of 5d occupancy considered above for the lanthanides. The ground-state calculations do not allow us to estimate the sensitivity of actinide mobility to the 5d configuration responsible for the electronic state chromatography effect for the metastable states. This would require interaction potential calculations for the excited metastable states. Experience with the lanthanide family shows that the present {\it ab initio} methods are likely applicable only for Ac$^+$ and Lr$^+$ ions in their 6d7s metastable states.

\subsection{Ionic radii}
\label{ssec:Acrad}

The models used in sec.~\ref{ssec:Larad}  for lanthanide ionic radii can also be tested for actinide ions. The results are summarized in figure~\ref{fig:Acrad} that follows the format of figure~\ref{fig:Larad}. Parameters of the electron distributions of the bare ions taken from the same source~\citep{Indelicato2007} split into two parallel trend lines for ions with the 7s$^2$ (lower) and 7s (upper) outer shell configurations. The WB radii for ion-He interactions available from the present calculations follow the opposite order. The obvious reason already discussed is the enhancement of repulsive electronic interactions for the filled outer s shell. The radii extracted from the mobility analysis within the HS model show qualitatively similar variations and agree well in magnitude with the results for the lanthanide ions. Figure~\ref{fig:Acrad} confirms \hl{that the results of HS model are much less consistent and informative in the case of Ar}. The 7s and 7s$^2$ trends are less evident for the parameters derived from interaction potentials and mobilities, except for the room-temperature HS result. The latter, however, wrongly predicts a  general increase of ionic radii with $Z$. It is important to mention in this regard the relative measurements of the drift times for the Pu$^+$-Am$^+$ and Cf$^+$-Fm$^+$ pairs of ions in Ar~\citep{Sewtz2003,Backe2005}. The HS model estimated the relative contraction of the ionic radii in these pairs as 3.1$\pm$1.3 and 2\%, respectively.~\cite{Indelicato2007}  have already discussed these variations in terms of electronic structure parameters of the bare ions. The present analysis indicates that the drift times in Ar at room temperature correspond to the mobility minimum and may not be sensitive to the effective size of an ion. Quantitative interpretation of such data within the oversimplified HS model requires caution, as pointed out by~\cite{Backe2015}.

\begin{figure}[hbt!]
\includegraphics[width=1.0\linewidth]{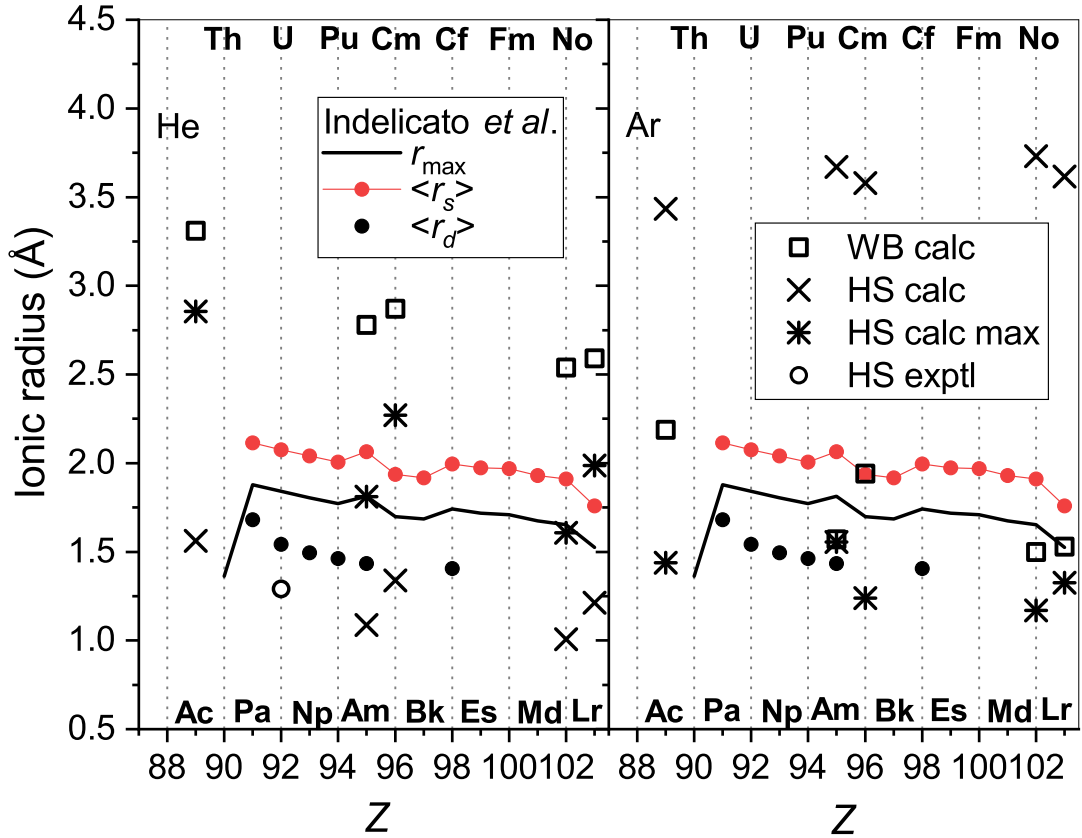}
\caption{The radii of the actinide ions determined from He (left panel) and Ar (right panel) data. Presented are WB radii from {\it ab initio} calculations and the results of the HS model applied to experimental and calculated room-temperature mobilities \hl{``HS exptl'' and ``HS calc'', respectively)} and to the calculated mobility at the maximum \hl{(``HS calc max'')}. Parameters of the ion electron distribution calculated by~\cite{Indelicato2007} are also shown; see sec.~\ref{ssec:Larad} for explanation.}
\label{fig:Acrad}
\end{figure}

\section{Conclusions and outlook}
\label{sec:conclude}

Progress in the one atom at a time production of the heavy and superheavy elements calls for new experimental techniques capable of characterizing the electronic structure of nascent or neutralized fusion products. Measurements of transport properties of the ions, in particular their gaseous mobilities, have already been counted among the most likely approaches, at least from the technical standpoint~\citep{Backe2015,Rickert2020}. The lanthanide and actinide families serve as a natural example for interpretation of such measurements in terms of electronic structure parameters. They can also provide solid grounds for assessing the accuracy of {\it ab initio} calculations of ion-atom interaction potentials and transport properties, which are invaluable for guiding complicated and expensive on-line experiments in one atom at a time mode.

The conclusion of the present analysis is that the mobility is very sensitive to the electronic configuration of the ion. Both room-temperature measurements and {\it ab initio} theoretical calculations for the lanthanide ions reveal sharp deviations in the mobilities of the 5d6s and 6s$^2$ ions from the trend line for the 6s ions, and slowly varying changes with 4f shell occupancy (equivalently, atomic number). Comparison between experiment and theory shows that the latter is presently able to predict the mobility differences for lanthanide ions in the ground and metastable states and to determine the conditions (buffer gas temperature, reduced electric field strength, pressure, etc.) for achieving the best discrimination of the ions by their drift times. Here, we have extended this conclusion to the actinides, which are unexplored experimentally.  We found significant difference in the mobility of 7s and 7s$^2$ ions, which finds qualitative confirmations in the spatial electron density distributions of the bare ions~\citep{Indelicato2007}. Supplementing the profound effect of the $n$d$^{m}$ $\leftrightarrow$ $n$d$^{m-1}(n+1)$s electron promotion on the mobility already known for transition metal ions~\citep{Kemper1991, Bowers1993, Taylor1999, Iceman2007, Ibrahim2008, Manard2016b, Manard2016a}, discrimination of the $n$s, $(n-1)$d$n$s and $n$s$^2$ configurations have direct implications for probing the electronic configurations of the superheavy elements with $Z=104-112$. Another important application is the so-called laser resonance chromatography proposal for indirect detection of the spectroscopic transitions by discrimination of the ions in metastable states~\citep{LRC2019}, which is currently being explored for the Lu$^+$ and Lr$^+$ ions.

The present overview demonstrates that the current theoretical state of the art allows one to interpret and predict trends in the mobility of heavy ions. Standard (and relatively cheap) scalar relativistic, single-reference, {\it ab initio} methods are able to link the electronic structure of selected ions and their transport properties by means of the ion-atom interaction potentials. Predicted changes in the mobility upon the electronic excitations are useful for advancing experimental methods of ion discrimination.  At the same time, the lack of experimental data strongly limits the quantitative assessment of the {\it ab initio} results and further development of the theory. Measurements of the mobility as function of temperature or $E/N$ are absent for most of the elements above Ba. Indeed, only two room-temperature mobility values for lanthanide ions, i.e. for Gd$^+$($^{10}$D) ion in He and Ar~\citep{Mustapha2012, Manard2017}, are available to compare the performance of the single- and multi-reference methods, to assess the role of vectorial spin-orbit coupling and to establish the uncertainty of the transport calculations for heavy, open-shell ions. Careful analysis reported here roughly estimated the respective variances as 20, 10 and 5\%~\cite{IJMS2}. This indicates the need for testing multi-reference coupled cluster techniques~\cite{MRCC} in combination with SO configuration interaction methods. Such a demanding approach, however, will probably need to be customized for each particular lanthanide ion for which experimental data exists. Experimental mobilities for the actinides and even more difficult theoretical calculations remain for the future. Only by means of aligned experimental and theoretical efforts can the frontier of ion transport studies be pushed from the present scattered reconnaissance to a legitimate inventory of heavy- and superheavy ion research.

%While the qualitative trends are evident, the lack of experimental data should be emphasized. Quantitative measurements of the mobility as function of temperature or electric field strength are especially in demand. For lanthanide ions, they would quantify the uncertainties of the theoretical modeling, not only those due to uncertainties in the interaction potentials but also those related to transport calculations for open-shell ions. For actinide ions, they should permit a direct test of the \hl{accuracy} of {\it ab initio} techniques, in particular, the role of relativistic effects \hl{and the validity of the scalar relativistic approximation}. On the theoretical side, accurate {\it ab initio} calculations of interaction potentials for ions with significant configuration mixing effects remain a major challenge. Future progress in these areas will be helpful for pushing the current frontier of ion transport studies from the present scattered reconnaissance to a legitimate inventory of heavy- and superheavy ion research.

\section*{Conflict of Interest Statement}
The authors declare that the research was conducted in the absence of any commercial or financial relationships that could be construed as a potential conflict of interest.

\section*{Author Contributions}
GV and AAB carried out the {\it ab initio} calculations and data analysis, LAV performed the calculations of ion mobilities and other transport properties, ML made a problem statement and assessment of the results and their implications to heavy and superheavy ion research. All athours contributed to the manuscript preparation.

\section*{Funding}
This work was supported by the Russian Foundation for Basic Research under the project No. 19-03-00144. M. Laatiaoui acknowledges funding from the European Research Council (ERC) under the European Union's Horizon 2020 research and innovation programme (grant agreement No. 819957).

\section*{Acknowledgments}
We thank Ms. Nika Buchachenko for her help with the manuscript formatting. Calculations were performed at Pardus and Arkuda Skoltech HPC clusters ({\it ab initio}) and at Chatham University (transport).

%\section*{Supplemental Data}
% \href{http://home.frontiersin.org/about/author-guidelines#SupplementaryMaterial}{Supplementary Material} should be uploaded separately on submission, if there are Supplementary Figures, please include the caption in the same file as the figure. LaTeX Supplementary Material templates can be found in the Frontiers LaTeX folder.

\section*{Data Availability Statement}
Interaction potentials, ion mobilities and other transport data are available in the~\cite{LXCAT} database within the LXCat Plasma Data Exchange Project [http://www.lxcat.net/Viehland, last accessed Feb. 11, 2020]. The transport data for various isotopes was obtained from the present results using the aliasing technique of~\cite{Viehland2016}.
% Please see the availability of data guidelines for more information, at https://www.frontiersin.org/about/author-guidelines#AvailabilityofData

%\bibliographystyle{frontiersinSCNS_ENG_HUMS} % for Science, Engineering and Humanities and Social Sciences articles, for Humanities and Social Sciences articles please include page numbers in the in-text citations
%\bibliographystyle{frontiersinHLTH&FPHY} % for Health, Physics and Mathematics articles
%\bibliographystyle{agsm}
\bibliography{fc_mobility_arxiv}

\begin{thebibliography}{79}%
\makeatletter
\providecommand \@ifxundefined [1]{%
 \@ifx{#1\undefined}
}%
\providecommand \@ifnum [1]{%
 \ifnum #1\expandafter \@firstoftwo
 \else \expandafter \@secondoftwo
 \fi
}%
\providecommand \@ifx [1]{%
 \ifx #1\expandafter \@firstoftwo
 \else \expandafter \@secondoftwo
 \fi
}%
\providecommand \natexlab [1]{#1}%
\providecommand \enquote  [1]{``#1''}%
\providecommand \bibnamefont  [1]{#1}%
\providecommand \bibfnamefont [1]{#1}%
\providecommand \citenamefont [1]{#1}%
\providecommand \href@noop [0]{\@secondoftwo}%
\providecommand \href [0]{\begingroup \@sanitize@url \@href}%
\providecommand \@href[1]{\@@startlink{#1}\@@href}%
\providecommand \@@href[1]{\endgroup#1\@@endlink}%
\providecommand \@sanitize@url [0]{\catcode `\\12\catcode `\$12\catcode
  `\&12\catcode `\#12\catcode `\^12\catcode `\_12\catcode `\%12\relax}%
\providecommand \@@startlink[1]{}%
\providecommand \@@endlink[0]{}%
\providecommand \url  [0]{\begingroup\@sanitize@url \@url }%
\providecommand \@url [1]{\endgroup\@href {#1}{\urlprefix }}%
\providecommand \urlprefix  [0]{URL }%
\providecommand \Eprint [0]{\href }%
\providecommand \doibase [0]{http://dx.doi.org/}%
\providecommand \selectlanguage [0]{\@gobble}%
\providecommand \bibinfo  [0]{\@secondoftwo}%
\providecommand \bibfield  [0]{\@secondoftwo}%
\providecommand \translation [1]{[#1]}%
\providecommand \BibitemOpen [0]{}%
\providecommand \bibitemStop [0]{}%
\providecommand \bibitemNoStop [0]{.\EOS\space}%
\providecommand \EOS [0]{\spacefactor3000\relax}%
\providecommand \BibitemShut  [1]{\csname bibitem#1\endcsname}%
\let\auto@bib@innerbib\@empty
%</preamble>
\bibitem [{\citenamefont {Karol}\ \emph
  {et~al.}(2016{\natexlab{a}})\citenamefont {Karol}, \citenamefont {Barber},
  \citenamefont {Sherrill}, \citenamefont {Vardaci},\ and\ \citenamefont
  {Yamazaki}}]{Karol2016b}%
  \BibitemOpen
  \bibfield  {author} {\bibinfo {author} {\bibfnamefont {P.~J.}\ \bibnamefont
  {Karol}}, \bibinfo {author} {\bibfnamefont {R.~C.}\ \bibnamefont {Barber}},
  \bibinfo {author} {\bibfnamefont {B.~M.}\ \bibnamefont {Sherrill}}, \bibinfo
  {author} {\bibfnamefont {E.}~\bibnamefont {Vardaci}}, \ and\ \bibinfo
  {author} {\bibfnamefont {T.}~\bibnamefont {Yamazaki}},\ }\href {\doibase
  10.1515/pac-2015-0501} {\bibfield  {journal} {\bibinfo  {journal} {Pure Appl.
  Chem.}\ }\textbf {\bibinfo {volume} {88}},\ \bibinfo {pages} {155} (\bibinfo
  {year} {2016}{\natexlab{a}})}\BibitemShut {NoStop}%
\bibitem [{\citenamefont {Karol}\ \emph
  {et~al.}(2016{\natexlab{b}})\citenamefont {Karol}, \citenamefont {Barber},
  \citenamefont {Sherrill}, \citenamefont {Vardaci},\ and\ \citenamefont
  {Yamazaki}}]{Karol2016a}%
  \BibitemOpen
  \bibfield  {author} {\bibinfo {author} {\bibfnamefont {P.~J.}\ \bibnamefont
  {Karol}}, \bibinfo {author} {\bibfnamefont {R.~C.}\ \bibnamefont {Barber}},
  \bibinfo {author} {\bibfnamefont {B.~M.}\ \bibnamefont {Sherrill}}, \bibinfo
  {author} {\bibfnamefont {E.}~\bibnamefont {Vardaci}}, \ and\ \bibinfo
  {author} {\bibfnamefont {T.}~\bibnamefont {Yamazaki}},\ }\href {\doibase
  10.1515/pac-2015-0502} {\bibfield  {journal} {\bibinfo  {journal} {Pure Appl.
  Chem.}\ }\textbf {\bibinfo {volume} {88}},\ \bibinfo {pages} {139} (\bibinfo
  {year} {2016}{\natexlab{b}})}\BibitemShut {NoStop}%
\bibitem [{\citenamefont {Perey}(1939)}]{Perey1939}%
  \BibitemOpen
  \bibfield  {author} {\bibinfo {author} {\bibfnamefont {M.}~\bibnamefont
  {Perey}},\ }\href {\doibase 10.1051/jphysrad:019390010010043500} {\bibfield
  {journal} {\bibinfo  {journal} {J. Phys. Radium}\ }\textbf {\bibinfo {volume}
  {10}},\ \bibinfo {pages} {435} (\bibinfo {year} {1939})}\BibitemShut
  {NoStop}%
\bibitem [{\citenamefont {Perrier}\ and\ \citenamefont
  {Segr{\`e}}(1947)}]{Perrier1947}%
  \BibitemOpen
  \bibfield  {author} {\bibinfo {author} {\bibfnamefont {C.}~\bibnamefont
  {Perrier}}\ and\ \bibinfo {author} {\bibfnamefont {E.}~\bibnamefont
  {Segr{\`e}}},\ }\href {\doibase 10.1038/159024a0} {\bibfield  {journal}
  {\bibinfo  {journal} {Nature}\ }\textbf {\bibinfo {volume} {159}},\ \bibinfo
  {pages} {24} (\bibinfo {year} {1947})}\BibitemShut {NoStop}%
\bibitem [{\citenamefont {Ghiorso}\ \emph
  {et~al.}(1955{\natexlab{a}})\citenamefont {Ghiorso}, \citenamefont {Harvey},
  \citenamefont {Choppin}, \citenamefont {Thompson},\ and\ \citenamefont
  {Seaborg}}]{Ghiorso1955b}%
  \BibitemOpen
  \bibfield  {author} {\bibinfo {author} {\bibfnamefont {A.}~\bibnamefont
  {Ghiorso}}, \bibinfo {author} {\bibfnamefont {B.~G.}\ \bibnamefont {Harvey}},
  \bibinfo {author} {\bibfnamefont {G.~R.}\ \bibnamefont {Choppin}}, \bibinfo
  {author} {\bibfnamefont {S.~G.}\ \bibnamefont {Thompson}}, \ and\ \bibinfo
  {author} {\bibfnamefont {G.~T.}\ \bibnamefont {Seaborg}},\ }\href {\doibase
  10.1103/PhysRev.98.1518} {\bibfield  {journal} {\bibinfo  {journal} {Phys.
  Rev.}\ }\textbf {\bibinfo {volume} {98}},\ \bibinfo {pages} {1518} (\bibinfo
  {year} {1955}{\natexlab{a}})}\BibitemShut {NoStop}%
\bibitem [{\citenamefont {Ghiorso}\ \emph
  {et~al.}(1955{\natexlab{b}})\citenamefont {Ghiorso}, \citenamefont
  {Thompson}, \citenamefont {Higgins}, \citenamefont {Seaborg}, \citenamefont
  {Studier}, \citenamefont {Fields}, \citenamefont {Diamond}, \citenamefont
  {Mech}, \citenamefont {Pyle}, \citenamefont {Huizenga}, \citenamefont
  {Hirsch}, \citenamefont {Manning}, \citenamefont {Browne}, \citenamefont
  {Smith},\ and\ \citenamefont {Spence}}]{Ghiorso1955a}%
  \BibitemOpen
  \bibfield  {author} {\bibinfo {author} {\bibfnamefont {A.}~\bibnamefont
  {Ghiorso}}, \bibinfo {author} {\bibfnamefont {S.~G.}\ \bibnamefont
  {Thompson}}, \bibinfo {author} {\bibfnamefont {G.~H.}\ \bibnamefont
  {Higgins}}, \bibinfo {author} {\bibfnamefont {G.~T.}\ \bibnamefont
  {Seaborg}}, \bibinfo {author} {\bibfnamefont {M.~H.}\ \bibnamefont
  {Studier}}, \bibinfo {author} {\bibfnamefont {P.~R.}\ \bibnamefont {Fields}},
  \bibinfo {author} {\bibfnamefont {H.}~\bibnamefont {Diamond}}, \bibinfo
  {author} {\bibfnamefont {J.~F.}\ \bibnamefont {Mech}}, \bibinfo {author}
  {\bibfnamefont {G.~L.}\ \bibnamefont {Pyle}}, \bibinfo {author}
  {\bibfnamefont {J.~R.}\ \bibnamefont {Huizenga}}, \bibinfo {author}
  {\bibfnamefont {A.}~\bibnamefont {Hirsch}}, \bibinfo {author} {\bibfnamefont
  {W.~M.}\ \bibnamefont {Manning}}, \bibinfo {author} {\bibfnamefont {C.~I.}\
  \bibnamefont {Browne}}, \bibinfo {author} {\bibfnamefont {H.~L.}\
  \bibnamefont {Smith}}, \ and\ \bibinfo {author} {\bibfnamefont {R.~W.}\
  \bibnamefont {Spence}},\ }\href {\doibase 10.1103/PhysRev.99.1048} {\bibfield
   {journal} {\bibinfo  {journal} {Phys. Rev.}\ }\textbf {\bibinfo {volume}
  {99}},\ \bibinfo {pages} {1048} (\bibinfo {year}
  {1955}{\natexlab{b}})}\BibitemShut {NoStop}%
\bibitem [{\citenamefont {T{\"u}rler}\ and\ \citenamefont
  {Pershina}(2013)}]{Turler2013}%
  \BibitemOpen
  \bibfield  {author} {\bibinfo {author} {\bibfnamefont {A.}~\bibnamefont
  {T{\"u}rler}}\ and\ \bibinfo {author} {\bibfnamefont {V.~G.}\ \bibnamefont
  {Pershina}},\ }\href {\doibase 10.1021/cr3002438} {\bibfield  {journal}
  {\bibinfo  {journal} {Chem. Rev.}\ }\textbf {\bibinfo {volume} {113}},\
  \bibinfo {pages} {1237} (\bibinfo {year} {2013})}\BibitemShut {NoStop}%
\bibitem [{\citenamefont {Haba}(2019)}]{Haba2019}%
  \BibitemOpen
  \bibfield  {author} {\bibinfo {author} {\bibfnamefont {H.}~\bibnamefont
  {Haba}},\ }\href {\doibase 10.1038/s41557-018-0191-8} {\bibfield  {journal}
  {\bibinfo  {journal} {Nature Chem.}\ }\textbf {\bibinfo {volume} {11}},\
  \bibinfo {pages} {10} (\bibinfo {year} {2019})}\BibitemShut {NoStop}%
\bibitem [{\citenamefont {Seaborg}(1945)}]{Seaborg1945}%
  \BibitemOpen
  \bibfield  {author} {\bibinfo {author} {\bibfnamefont {G.~T.}\ \bibnamefont
  {Seaborg}},\ }\href {\doibase 10.1021/cen-v023n023.p2190} {\bibfield
  {journal} {\bibinfo  {journal} {Chem. Eng. News}\ }\textbf {\bibinfo {volume}
  {23}},\ \bibinfo {pages} {2190} (\bibinfo {year} {1945})}\BibitemShut
  {NoStop}%
\bibitem [{\citenamefont {Seaborg}\ and\ \citenamefont
  {Loveland}(1990)}]{Seaborg1990}%
  \BibitemOpen
  \bibfield  {author} {\bibinfo {author} {\bibfnamefont {G.~T.}\ \bibnamefont
  {Seaborg}}\ and\ \bibinfo {author} {\bibfnamefont {W.~D.}\ \bibnamefont
  {Loveland}},\ }\href@noop {} {\emph {\bibinfo {title} {The Elements Beyond
  Uranium}}}\ (\bibinfo  {publisher} {J. Wiley and Sons},\ \bibinfo {address}
  {New York},\ \bibinfo {year} {1990})\BibitemShut {NoStop}%
\bibitem [{\citenamefont {Wallmann}(1959)}]{Wallmann1959}%
  \BibitemOpen
  \bibfield  {author} {\bibinfo {author} {\bibfnamefont {J.~C.}\ \bibnamefont
  {Wallmann}},\ }\href {\doibase 10.1021/ed036p340} {\bibfield  {journal}
  {\bibinfo  {journal} {J. Chem. Educ.}\ }\textbf {\bibinfo {volume} {36}},\
  \bibinfo {pages} {340} (\bibinfo {year} {1959})}\BibitemShut {NoStop}%
\bibitem [{\citenamefont {Eds.~Sch{\"a}del}\ and\ \citenamefont
  {Shaughnessy}(2014)}]{SchadelBook2014}%
  \BibitemOpen
  \bibfield  {author} {\bibinfo {author} {\bibfnamefont {M.}~\bibnamefont
  {Eds.~Sch{\"a}del}}\ and\ \bibinfo {author} {\bibfnamefont {D.}~\bibnamefont
  {Shaughnessy}},\ }\href@noop {} {\emph {\bibinfo {title} {The Chemistry of
  Superheavy Elements}}}\ (\bibinfo  {publisher} {Springer},\ \bibinfo
  {address} {Berlin, Heidelberg},\ \bibinfo {year} {2014})\BibitemShut
  {NoStop}%
\bibitem [{\citenamefont {Oganessian}\ and\ \citenamefont
  {Dmitriev}(2016)}]{Oganessian2016}%
  \BibitemOpen
  \bibfield  {author} {\bibinfo {author} {\bibfnamefont {Y.~T.}\ \bibnamefont
  {Oganessian}}\ and\ \bibinfo {author} {\bibfnamefont {S.~N.}\ \bibnamefont
  {Dmitriev}},\ }\href {\doibase 10.1070/RCR4607} {\bibfield  {journal}
  {\bibinfo  {journal} {Russ. Chem. Rev.}\ }\textbf {\bibinfo {volume} {85}},\
  \bibinfo {pages} {901} (\bibinfo {year} {2016})}\BibitemShut {NoStop}%
\bibitem [{\citenamefont {Eichler}(2017)}]{Eichler2017}%
  \BibitemOpen
  \bibfield  {author} {\bibinfo {author} {\bibfnamefont {R.}~\bibnamefont
  {Eichler}},\ }in\ \href {\doibase 10.1007/978-3-319-44165-8_4} {\emph
  {\bibinfo {booktitle} {New Horizons in Fundamental Physics, FIAS
  Interdisciplinary Science Series}}},\ \bibinfo {editor} {edited by\ \bibinfo
  {editor} {\bibfnamefont {S.}~\bibnamefont {Schramm}}\ and\ \bibinfo {editor}
  {\bibfnamefont {M.}~\bibnamefont {Schafer}}}\ (\bibinfo  {publisher}
  {Springer},\ \bibinfo {address} {Cham},\ \bibinfo {year} {2017})\ pp.\
  \bibinfo {pages} {41--53}\BibitemShut {NoStop}%
\bibitem [{\citenamefont {Eichler}(2019)}]{Eichler2019}%
  \BibitemOpen
  \bibfield  {author} {\bibinfo {author} {\bibfnamefont {R.}~\bibnamefont
  {Eichler}},\ }\href {\doibase 10.1515/ract-2018-3080} {\bibfield  {journal}
  {\bibinfo  {journal} {Radiochim. Acta}\ }\textbf {\bibinfo {volume} {107}},\
  \bibinfo {pages} {865} (\bibinfo {year} {2019})}\BibitemShut {NoStop}%
\bibitem [{\citenamefont {D{\"u}llmann}(2019)}]{Duellmann2019}%
  \BibitemOpen
  \bibfield  {author} {\bibinfo {author} {\bibfnamefont {C.~E.}\ \bibnamefont
  {D{\"u}llmann}},\ }\href {\doibase 10.1515/ract-2019-0012} {\bibfield
  {journal} {\bibinfo  {journal} {Radiochim Acta}\ }\textbf {\bibinfo {volume}
  {107}},\ \bibinfo {pages} {587} (\bibinfo {year} {2019})}\BibitemShut
  {NoStop}%
\bibitem [{\citenamefont {Ter-Akopian}\ and\ \citenamefont
  {Dmitriev}(2015)}]{TerAkopian2015}%
  \BibitemOpen
  \bibfield  {author} {\bibinfo {author} {\bibfnamefont {G.~M.}\ \bibnamefont
  {Ter-Akopian}}\ and\ \bibinfo {author} {\bibfnamefont {S.~N.}\ \bibnamefont
  {Dmitriev}},\ }\href {\doibase 10.1016/j.nuclphysa.2015.09.004} {\bibfield
  {journal} {\bibinfo  {journal} {Nucl. Phys. A}\ }\textbf {\bibinfo {volume}
  {944}},\ \bibinfo {pages} {177} (\bibinfo {year} {2015})}\BibitemShut
  {NoStop}%
\bibitem [{\citenamefont {Pershina}(1996)}]{Pershina1996}%
  \BibitemOpen
  \bibfield  {author} {\bibinfo {author} {\bibfnamefont {V.~G.}\ \bibnamefont
  {Pershina}},\ }\href {\doibase 10.1021/cr941182g} {\bibfield  {journal}
  {\bibinfo  {journal} {Chem. Rev.}\ }\textbf {\bibinfo {volume} {96}},\
  \bibinfo {pages} {1977} (\bibinfo {year} {1996})}\BibitemShut {NoStop}%
\bibitem [{\citenamefont {Pyykk{\"o}}(2012)}]{Pyykko2012}%
  \BibitemOpen
  \bibfield  {author} {\bibinfo {author} {\bibfnamefont {P.}~\bibnamefont
  {Pyykk{\"o}}},\ }\href {\doibase 10.1021/cr200042e} {\bibfield  {journal}
  {\bibinfo  {journal} {Chem. Rev.}\ }\textbf {\bibinfo {volume} {112}},\
  \bibinfo {pages} {371} (\bibinfo {year} {2012})}\BibitemShut {NoStop}%
\bibitem [{\citenamefont {Eliav}\ \emph {et~al.}(2015)\citenamefont {Eliav},
  \citenamefont {Fritzsche},\ and\ \citenamefont {Kaldor}}]{Eliav2015}%
  \BibitemOpen
  \bibfield  {author} {\bibinfo {author} {\bibfnamefont {E.}~\bibnamefont
  {Eliav}}, \bibinfo {author} {\bibfnamefont {S.}~\bibnamefont {Fritzsche}}, \
  and\ \bibinfo {author} {\bibfnamefont {U.}~\bibnamefont {Kaldor}},\ }\href
  {\doibase 10.1016/j.nuclphysa.2015.06.017} {\bibfield  {journal} {\bibinfo
  {journal} {Nucl. Phys. A}\ }\textbf {\bibinfo {volume} {944}},\ \bibinfo
  {pages} {518} (\bibinfo {year} {2015})}\BibitemShut {NoStop}%
\bibitem [{\citenamefont {Pyykk{\"o}}(2016)}]{Pyykko2016}%
  \BibitemOpen
  \bibfield  {author} {\bibinfo {author} {\bibfnamefont {P.}~\bibnamefont
  {Pyykk{\"o}}},\ }\href {\doibase 10.1051/epjconf/201613101001} {\bibfield
  {journal} {\bibinfo  {journal} {Eur. Phys. J. Web Conf.}\ }\textbf {\bibinfo
  {volume} {131}},\ \bibinfo {pages} {01001} (\bibinfo {year}
  {2016})}\BibitemShut {NoStop}%
\bibitem [{\citenamefont {Ed.~Liu}(2017)}]{LiuBook2017}%
  \BibitemOpen
  \bibfield  {author} {\bibinfo {author} {\bibfnamefont {W.}~\bibnamefont
  {Ed.~Liu}},\ }\href@noop {} {\emph {\bibinfo {title} {Handbook of
  Relativistic Quantum Chemistry}}}\ (\bibinfo  {publisher} {Springer},\
  \bibinfo {address} {Berlin},\ \bibinfo {year} {2017})\BibitemShut {NoStop}%
\bibitem [{\citenamefont {Dzuba}\ \emph {et~al.}(2017)\citenamefont {Dzuba},
  \citenamefont {Flambaum},\ and\ \citenamefont {Webb}}]{Dzuba2017}%
  \BibitemOpen
  \bibfield  {author} {\bibinfo {author} {\bibfnamefont {V.~A.}\ \bibnamefont
  {Dzuba}}, \bibinfo {author} {\bibfnamefont {V.~V.}\ \bibnamefont {Flambaum}},
  \ and\ \bibinfo {author} {\bibfnamefont {J.~K.}\ \bibnamefont {Webb}},\
  }\href {\doibase 10.1103/PhysRevA.95.062515} {\bibfield  {journal} {\bibinfo
  {journal} {Phys. Rev. A}\ }\textbf {\bibinfo {volume} {95}},\ \bibinfo
  {pages} {062515} (\bibinfo {year} {2017})}\BibitemShut {NoStop}%
\bibitem [{\citenamefont {Giuliani}\ \emph {et~al.}(2019)\citenamefont
  {Giuliani}, \citenamefont {Matheson}, \citenamefont {Nazarewicz},
  \citenamefont {Olsen}, \citenamefont {Reinhard}, \citenamefont {Sadhukhan},
  \citenamefont {Schuetrumpf}, \citenamefont {Schunck},\ and\ \citenamefont
  {Schwerdtfeger}}]{Giuliani2019}%
  \BibitemOpen
  \bibfield  {author} {\bibinfo {author} {\bibfnamefont {S.~A.}\ \bibnamefont
  {Giuliani}}, \bibinfo {author} {\bibfnamefont {Z.}~\bibnamefont {Matheson}},
  \bibinfo {author} {\bibfnamefont {W.}~\bibnamefont {Nazarewicz}}, \bibinfo
  {author} {\bibfnamefont {E.}~\bibnamefont {Olsen}}, \bibinfo {author}
  {\bibfnamefont {P.~G.}\ \bibnamefont {Reinhard}}, \bibinfo {author}
  {\bibfnamefont {J.}~\bibnamefont {Sadhukhan}}, \bibinfo {author}
  {\bibfnamefont {B.}~\bibnamefont {Schuetrumpf}}, \bibinfo {author}
  {\bibfnamefont {N.}~\bibnamefont {Schunck}}, \ and\ \bibinfo {author}
  {\bibfnamefont {P.}~\bibnamefont {Schwerdtfeger}},\ }\href {\doibase
  10.1103/RevModPhys.91.011001} {\bibfield  {journal} {\bibinfo  {journal}
  {Rev. Mod. Phys.}\ }\textbf {\bibinfo {volume} {91}},\ \bibinfo {pages}
  {011001} (\bibinfo {year} {2019})}\BibitemShut {NoStop}%
\bibitem [{\citenamefont {Backe}\ \emph {et~al.}(2015)\citenamefont {Backe},
  \citenamefont {Lauth}, \citenamefont {Block},\ and\ \citenamefont
  {Laatiaoui}}]{Backe2015}%
  \BibitemOpen
  \bibfield  {author} {\bibinfo {author} {\bibfnamefont {H.}~\bibnamefont
  {Backe}}, \bibinfo {author} {\bibfnamefont {W.}~\bibnamefont {Lauth}},
  \bibinfo {author} {\bibfnamefont {M.}~\bibnamefont {Block}}, \ and\ \bibinfo
  {author} {\bibfnamefont {M.}~\bibnamefont {Laatiaoui}},\ }\href {\doibase
  10.1016/j.nuclphysa.2015.07.002} {\bibfield  {journal} {\bibinfo  {journal}
  {Nucl. Phys. A}\ }\textbf {\bibinfo {volume} {944}},\ \bibinfo {pages} {492}
  (\bibinfo {year} {2015})}\BibitemShut {NoStop}%
\bibitem [{\citenamefont {Sewtz}\ \emph {et~al.}(2003)\citenamefont {Sewtz},
  \citenamefont {Backe}, \citenamefont {Dretzke}, \citenamefont {Kube},
  \citenamefont {Lauth}, \citenamefont {Schwamb}, \citenamefont {Eberhardt},
  \citenamefont {Gruning}, \citenamefont {Th{\"o}rle}, \citenamefont
  {Trautmann}, \citenamefont {Kunz}, \citenamefont {Lassen}, \citenamefont
  {Passler}, \citenamefont {Dong}, \citenamefont {Fritzsche},\ and\
  \citenamefont {Haire}}]{Sewtz2003}%
  \BibitemOpen
  \bibfield  {author} {\bibinfo {author} {\bibfnamefont {M.}~\bibnamefont
  {Sewtz}}, \bibinfo {author} {\bibfnamefont {H.}~\bibnamefont {Backe}},
  \bibinfo {author} {\bibfnamefont {A.}~\bibnamefont {Dretzke}}, \bibinfo
  {author} {\bibfnamefont {G.}~\bibnamefont {Kube}}, \bibinfo {author}
  {\bibfnamefont {W.}~\bibnamefont {Lauth}}, \bibinfo {author} {\bibfnamefont
  {P.}~\bibnamefont {Schwamb}}, \bibinfo {author} {\bibfnamefont
  {K.}~\bibnamefont {Eberhardt}}, \bibinfo {author} {\bibfnamefont
  {C.}~\bibnamefont {Gruning}}, \bibinfo {author} {\bibfnamefont
  {P.}~\bibnamefont {Th{\"o}rle}}, \bibinfo {author} {\bibfnamefont
  {N.}~\bibnamefont {Trautmann}}, \bibinfo {author} {\bibfnamefont
  {P.}~\bibnamefont {Kunz}}, \bibinfo {author} {\bibfnamefont {J.}~\bibnamefont
  {Lassen}}, \bibinfo {author} {\bibfnamefont {G.}~\bibnamefont {Passler}},
  \bibinfo {author} {\bibfnamefont {C.~Z.}\ \bibnamefont {Dong}}, \bibinfo
  {author} {\bibfnamefont {S.}~\bibnamefont {Fritzsche}}, \ and\ \bibinfo
  {author} {\bibfnamefont {R.~G.}\ \bibnamefont {Haire}},\ }\href {\doibase
  10.1103/PhysRevLett.90.163002} {\bibfield  {journal} {\bibinfo  {journal}
  {Phys. Rev. Lett.}\ }\textbf {\bibinfo {volume} {90}},\ \bibinfo {pages}
  {163002} (\bibinfo {year} {2003})}\BibitemShut {NoStop}%
\bibitem [{\citenamefont {Laatiaoui}\ \emph {et~al.}(2016)\citenamefont
  {Laatiaoui}, \citenamefont {Lauth}, \citenamefont {Backe}, \citenamefont
  {Block}, \citenamefont {Ackermann}, \citenamefont {Cheal}, \citenamefont
  {Chhetri}, \citenamefont {D{\"u}llmann}, \citenamefont {van Duppen},
  \citenamefont {Even}, \citenamefont {Ferrer}, \citenamefont {Giacoppo},
  \citenamefont {G{\"o}tz}, \citenamefont {He{\ss}berger}, \citenamefont
  {Huyse}, \citenamefont {Kaleja}, \citenamefont {Khuyagbaatar}, \citenamefont
  {Kunz}, \citenamefont {Lautenschl{\"a}ger}, \citenamefont {Mistry},
  \citenamefont {Raeder}, \citenamefont {Ramirez}, \citenamefont {Walther},
  \citenamefont {Wraith},\ and\ \citenamefont {Yakushev}}]{Laatiaoui2016}%
  \BibitemOpen
  \bibfield  {author} {\bibinfo {author} {\bibfnamefont {M.}~\bibnamefont
  {Laatiaoui}}, \bibinfo {author} {\bibfnamefont {W.}~\bibnamefont {Lauth}},
  \bibinfo {author} {\bibfnamefont {H.}~\bibnamefont {Backe}}, \bibinfo
  {author} {\bibfnamefont {M.}~\bibnamefont {Block}}, \bibinfo {author}
  {\bibfnamefont {D.}~\bibnamefont {Ackermann}}, \bibinfo {author}
  {\bibfnamefont {B.}~\bibnamefont {Cheal}}, \bibinfo {author} {\bibfnamefont
  {P.}~\bibnamefont {Chhetri}}, \bibinfo {author} {\bibfnamefont {C.~E.}\
  \bibnamefont {D{\"u}llmann}}, \bibinfo {author} {\bibfnamefont
  {P.}~\bibnamefont {van Duppen}}, \bibinfo {author} {\bibfnamefont
  {J.}~\bibnamefont {Even}}, \bibinfo {author} {\bibfnamefont {R.}~\bibnamefont
  {Ferrer}}, \bibinfo {author} {\bibfnamefont {F.}~\bibnamefont {Giacoppo}},
  \bibinfo {author} {\bibfnamefont {S.}~\bibnamefont {G{\"o}tz}}, \bibinfo
  {author} {\bibfnamefont {F.~P.}\ \bibnamefont {He{\ss}berger}}, \bibinfo
  {author} {\bibfnamefont {M.}~\bibnamefont {Huyse}}, \bibinfo {author}
  {\bibfnamefont {O.}~\bibnamefont {Kaleja}}, \bibinfo {author} {\bibfnamefont
  {J.}~\bibnamefont {Khuyagbaatar}}, \bibinfo {author} {\bibfnamefont
  {P.}~\bibnamefont {Kunz}}, \bibinfo {author} {\bibfnamefont {F.}~\bibnamefont
  {Lautenschl{\"a}ger}}, \bibinfo {author} {\bibfnamefont {A.~K.}\ \bibnamefont
  {Mistry}}, \bibinfo {author} {\bibfnamefont {S.}~\bibnamefont {Raeder}},
  \bibinfo {author} {\bibfnamefont {E.~M.}\ \bibnamefont {Ramirez}}, \bibinfo
  {author} {\bibfnamefont {T.}~\bibnamefont {Walther}}, \bibinfo {author}
  {\bibfnamefont {C.}~\bibnamefont {Wraith}}, \ and\ \bibinfo {author}
  {\bibfnamefont {A.}~\bibnamefont {Yakushev}},\ }\href {\doibase
  10.1038/nature19345} {\bibfield  {journal} {\bibinfo  {journal} {Nature}\
  }\textbf {\bibinfo {volume} {538}},\ \bibinfo {pages} {495} (\bibinfo {year}
  {2016})}\BibitemShut {NoStop}%
\bibitem [{\citenamefont {Chhetri}\ \emph {et~al.}(2018)\citenamefont
  {Chhetri}, \citenamefont {Ackermann}, \citenamefont {Backe}, \citenamefont
  {Block}, \citenamefont {Cheal}, \citenamefont {Droese}, \citenamefont
  {D{\"u}llmann}, \citenamefont {Even}, \citenamefont {Ferrer}, \citenamefont
  {Giacoppo}, \citenamefont {G{\"o}tz}, \citenamefont {He{\ss}berger},
  \citenamefont {Huyse}, \citenamefont {Kaleja}, \citenamefont {Khuyagbaatar},
  \citenamefont {Kunz}, \citenamefont {Laatiaoui}, \citenamefont
  {Lautenschl{\"a}ger}, \citenamefont {Lauth}, \citenamefont {Lecesne},
  \citenamefont {Lens}, \citenamefont {Ramirez}, \citenamefont {Mistry},
  \citenamefont {Raeder}, \citenamefont {van Duppen}, \citenamefont {Walther},
  \citenamefont {Yakushev},\ and\ \citenamefont {Zhang}}]{Chhetri2018}%
  \BibitemOpen
  \bibfield  {author} {\bibinfo {author} {\bibfnamefont {P.}~\bibnamefont
  {Chhetri}}, \bibinfo {author} {\bibfnamefont {D.}~\bibnamefont {Ackermann}},
  \bibinfo {author} {\bibfnamefont {H.}~\bibnamefont {Backe}}, \bibinfo
  {author} {\bibfnamefont {M.}~\bibnamefont {Block}}, \bibinfo {author}
  {\bibfnamefont {B.}~\bibnamefont {Cheal}}, \bibinfo {author} {\bibfnamefont
  {C.}~\bibnamefont {Droese}}, \bibinfo {author} {\bibfnamefont {C.~E.}\
  \bibnamefont {D{\"u}llmann}}, \bibinfo {author} {\bibfnamefont
  {J.}~\bibnamefont {Even}}, \bibinfo {author} {\bibfnamefont {R.}~\bibnamefont
  {Ferrer}}, \bibinfo {author} {\bibfnamefont {F.}~\bibnamefont {Giacoppo}},
  \bibinfo {author} {\bibfnamefont {S.}~\bibnamefont {G{\"o}tz}}, \bibinfo
  {author} {\bibfnamefont {F.~P.}\ \bibnamefont {He{\ss}berger}}, \bibinfo
  {author} {\bibfnamefont {M.}~\bibnamefont {Huyse}}, \bibinfo {author}
  {\bibfnamefont {O.}~\bibnamefont {Kaleja}}, \bibinfo {author} {\bibfnamefont
  {J.}~\bibnamefont {Khuyagbaatar}}, \bibinfo {author} {\bibfnamefont
  {P.}~\bibnamefont {Kunz}}, \bibinfo {author} {\bibfnamefont {M.}~\bibnamefont
  {Laatiaoui}}, \bibinfo {author} {\bibfnamefont {F.}~\bibnamefont
  {Lautenschl{\"a}ger}}, \bibinfo {author} {\bibfnamefont {W.}~\bibnamefont
  {Lauth}}, \bibinfo {author} {\bibfnamefont {N.}~\bibnamefont {Lecesne}},
  \bibinfo {author} {\bibfnamefont {L.}~\bibnamefont {Lens}}, \bibinfo {author}
  {\bibfnamefont {E.~M.}\ \bibnamefont {Ramirez}}, \bibinfo {author}
  {\bibfnamefont {A.~K.}\ \bibnamefont {Mistry}}, \bibinfo {author}
  {\bibfnamefont {S.}~\bibnamefont {Raeder}}, \bibinfo {author} {\bibfnamefont
  {P.}~\bibnamefont {van Duppen}}, \bibinfo {author} {\bibfnamefont
  {T.}~\bibnamefont {Walther}}, \bibinfo {author} {\bibfnamefont
  {A.}~\bibnamefont {Yakushev}}, \ and\ \bibinfo {author} {\bibfnamefont
  {Z.}~\bibnamefont {Zhang}},\ }\href {\doibase 10.1103/PhysRevLett.120.263003}
  {\bibfield  {journal} {\bibinfo  {journal} {Phys. Rev. Lett.}\ }\textbf
  {\bibinfo {volume} {120}},\ \bibinfo {pages} {263003} (\bibinfo {year}
  {2018})}\BibitemShut {NoStop}%
\bibitem [{\citenamefont {Campbell}\ \emph {et~al.}(2016)\citenamefont
  {Campbell}, \citenamefont {Moore},\ and\ \citenamefont
  {Pearson}}]{Campbell2016}%
  \BibitemOpen
  \bibfield  {author} {\bibinfo {author} {\bibfnamefont {P.}~\bibnamefont
  {Campbell}}, \bibinfo {author} {\bibfnamefont {I.~D.}\ \bibnamefont {Moore}},
  \ and\ \bibinfo {author} {\bibfnamefont {M.~R.}\ \bibnamefont {Pearson}},\
  }\href {\doibase 10.1016/j.ppnp.2015.09.003} {\bibfield  {journal} {\bibinfo
  {journal} {Progr. Part. Nucl. Phys.}\ }\textbf {\bibinfo {volume} {86}},\
  \bibinfo {pages} {127} (\bibinfo {year} {2016})}\BibitemShut {NoStop}%
\bibitem [{\citenamefont {Rickert}\ \emph {et~al.}(2020)\citenamefont
  {Rickert}, \citenamefont {Backe}, \citenamefont {Block}, \citenamefont
  {Laatiaoui}, \citenamefont {Lauth}, \citenamefont {Schneider},\ and\
  \citenamefont {Schneider}}]{Rickert2020}%
  \BibitemOpen
  \bibfield  {author} {\bibinfo {author} {\bibfnamefont {E.}~\bibnamefont
  {Rickert}}, \bibinfo {author} {\bibfnamefont {H.}~\bibnamefont {Backe}},
  \bibinfo {author} {\bibfnamefont {M.}~\bibnamefont {Block}}, \bibinfo
  {author} {\bibfnamefont {M.}~\bibnamefont {Laatiaoui}}, \bibinfo {author}
  {\bibfnamefont {W.}~\bibnamefont {Lauth}}, \bibinfo {author} {\bibfnamefont
  {J.}~\bibnamefont {Schneider}}, \ and\ \bibinfo {author} {\bibfnamefont
  {F.}~\bibnamefont {Schneider}},\ }\href {\doibase 10.1007/s10751-019-1691-7}
  {\bibfield  {journal} {\bibinfo  {journal} {Hyper. Int. (in press)}\ }
  (\bibinfo {year} {2020}),\ 10.1007/s10751-019-1691-7}\BibitemShut {NoStop}%
\bibitem [{\citenamefont {Kemper}\ and\ \citenamefont
  {Bowers}(1991)}]{Kemper1991}%
  \BibitemOpen
  \bibfield  {author} {\bibinfo {author} {\bibfnamefont {P.~R.}\ \bibnamefont
  {Kemper}}\ and\ \bibinfo {author} {\bibfnamefont {M.~T.}\ \bibnamefont
  {Bowers}},\ }\href {\doibase 10.1021/j100166a042} {\bibfield  {journal}
  {\bibinfo  {journal} {J. Phys. Chem.}\ }\textbf {\bibinfo {volume} {95}},\
  \bibinfo {pages} {5134} (\bibinfo {year} {1991})}\BibitemShut {NoStop}%
\bibitem [{\citenamefont {Bowers}\ \emph {et~al.}(1993)\citenamefont {Bowers},
  \citenamefont {Kemper}, \citenamefont {von Helden},\ and\ \citenamefont {van
  Koppen}}]{Bowers1993}%
  \BibitemOpen
  \bibfield  {author} {\bibinfo {author} {\bibfnamefont {M.~T.}\ \bibnamefont
  {Bowers}}, \bibinfo {author} {\bibfnamefont {P.~R.}\ \bibnamefont {Kemper}},
  \bibinfo {author} {\bibfnamefont {G.}~\bibnamefont {von Helden}}, \ and\
  \bibinfo {author} {\bibfnamefont {P.~A.~M.}\ \bibnamefont {van Koppen}},\
  }\href {\doibase 10.1126/science.260.5113.1446} {\bibfield  {journal}
  {\bibinfo  {journal} {Science}\ }\textbf {\bibinfo {volume} {260}},\ \bibinfo
  {pages} {1446} (\bibinfo {year} {1993})}\BibitemShut {NoStop}%
\bibitem [{\citenamefont {Taylor}\ \emph {et~al.}(1999)\citenamefont {Taylor},
  \citenamefont {Spicer},\ and\ \citenamefont {Barnas}}]{Taylor1999}%
  \BibitemOpen
  \bibfield  {author} {\bibinfo {author} {\bibfnamefont {W.~S.}\ \bibnamefont
  {Taylor}}, \bibinfo {author} {\bibfnamefont {E.~M.}\ \bibnamefont {Spicer}},
  \ and\ \bibinfo {author} {\bibfnamefont {D.~F.}\ \bibnamefont {Barnas}},\
  }\href {\doibase 10.1021/jp983887i} {\bibfield  {journal} {\bibinfo
  {journal} {J. Phys. Chem. A}\ }\textbf {\bibinfo {volume} {103}},\ \bibinfo
  {pages} {643} (\bibinfo {year} {1999})}\BibitemShut {NoStop}%
\bibitem [{\citenamefont {Iceman}\ \emph {et~al.}(2007)\citenamefont {Iceman},
  \citenamefont {Rue}, \citenamefont {Moision}, \citenamefont {Chatterjee},\
  and\ \citenamefont {Armentrout}}]{Iceman2007}%
  \BibitemOpen
  \bibfield  {author} {\bibinfo {author} {\bibfnamefont {C.}~\bibnamefont
  {Iceman}}, \bibinfo {author} {\bibfnamefont {C.}~\bibnamefont {Rue}},
  \bibinfo {author} {\bibfnamefont {R.~M.}\ \bibnamefont {Moision}}, \bibinfo
  {author} {\bibfnamefont {B.~K.}\ \bibnamefont {Chatterjee}}, \ and\ \bibinfo
  {author} {\bibfnamefont {P.~B.}\ \bibnamefont {Armentrout}},\ }\href
  {\doibase 10.1016/j.jasms.2007.02.012} {\bibfield  {journal} {\bibinfo
  {journal} {J. Am. Soc. Mass Spectrom.}\ }\textbf {\bibinfo {volume} {18}},\
  \bibinfo {pages} {1196} (\bibinfo {year} {2007})}\BibitemShut {NoStop}%
\bibitem [{\citenamefont {Ibrahim}\ \emph {et~al.}(2008)\citenamefont
  {Ibrahim}, \citenamefont {Alsharaeh}, \citenamefont {Mabrouki}, \citenamefont
  {Momoh}, \citenamefont {Xie},\ and\ \citenamefont {El-Shall}}]{Ibrahim2008}%
  \BibitemOpen
  \bibfield  {author} {\bibinfo {author} {\bibfnamefont {Y.}~\bibnamefont
  {Ibrahim}}, \bibinfo {author} {\bibfnamefont {E.}~\bibnamefont {Alsharaeh}},
  \bibinfo {author} {\bibfnamefont {R.}~\bibnamefont {Mabrouki}}, \bibinfo
  {author} {\bibfnamefont {P.}~\bibnamefont {Momoh}}, \bibinfo {author}
  {\bibfnamefont {E.}~\bibnamefont {Xie}}, \ and\ \bibinfo {author}
  {\bibfnamefont {M.~S.}\ \bibnamefont {El-Shall}},\ }\href {\doibase
  10.1021/jp077477i} {\bibfield  {journal} {\bibinfo  {journal} {J. Phys. Chem.
  A}\ }\textbf {\bibinfo {volume} {112}},\ \bibinfo {pages} {1112} (\bibinfo
  {year} {2008})}\BibitemShut {NoStop}%
\bibitem [{\citenamefont {Manard}\ and\ \citenamefont
  {Kemper}(2016{\natexlab{a}})}]{Manard2016b}%
  \BibitemOpen
  \bibfield  {author} {\bibinfo {author} {\bibfnamefont {M.~J.}\ \bibnamefont
  {Manard}}\ and\ \bibinfo {author} {\bibfnamefont {P.~R.}\ \bibnamefont
  {Kemper}},\ }\href {\doibase 10.1016/j.ijms.2016.07.006} {\bibfield
  {journal} {\bibinfo  {journal} {Int. J. Mass Spectrom.}\ }\textbf {\bibinfo
  {volume} {407}},\ \bibinfo {pages} {69} (\bibinfo {year}
  {2016}{\natexlab{a}})}\BibitemShut {NoStop}%
\bibitem [{\citenamefont {Manard}\ and\ \citenamefont
  {Kemper}(2016{\natexlab{b}})}]{Manard2016a}%
  \BibitemOpen
  \bibfield  {author} {\bibinfo {author} {\bibfnamefont {M.~J.}\ \bibnamefont
  {Manard}}\ and\ \bibinfo {author} {\bibfnamefont {P.~R.}\ \bibnamefont
  {Kemper}},\ }\href {\doibase 10.1016/j.ijms.2016.02.014} {\bibfield
  {journal} {\bibinfo  {journal} {Int. J. Mass Spectrom.}\ }\textbf {\bibinfo
  {volume} {402}},\ \bibinfo {pages} {1} (\bibinfo {year}
  {2016}{\natexlab{b}})}\BibitemShut {NoStop}%
\bibitem [{\citenamefont {Mason}\ and\ \citenamefont
  {McDaniel}(1988)}]{MasonBook1988}%
  \BibitemOpen
  \bibfield  {author} {\bibinfo {author} {\bibfnamefont {E.}~\bibnamefont
  {Mason}}\ and\ \bibinfo {author} {\bibfnamefont {E.}~\bibnamefont
  {McDaniel}},\ }\href@noop {} {\emph {\bibinfo {title} {Transport Properties
  of Ions in Gases}}}\ (\bibinfo  {publisher} {John Wiley and Sons},\ \bibinfo
  {address} {New York},\ \bibinfo {year} {1988})\BibitemShut {NoStop}%
\bibitem [{\citenamefont {Viehland}(2018)}]{LarryBook2018}%
  \BibitemOpen
  \bibfield  {author} {\bibinfo {author} {\bibfnamefont {L.~A.}\ \bibnamefont
  {Viehland}},\ }\href@noop {} {\emph {\bibinfo {title} {Gaseous Ion Mobility,
  Diffusion, and Reaction}}}\ (\bibinfo  {publisher} {Springer},\ \bibinfo
  {address} {Cham},\ \bibinfo {year} {2018})\BibitemShut {NoStop}%
\bibitem [{\citenamefont {Stone}(2013)}]{StoneBook}%
  \BibitemOpen
  \bibfield  {author} {\bibinfo {author} {\bibfnamefont {A.~J.}\ \bibnamefont
  {Stone}},\ }\href@noop {} {\emph {\bibinfo {title} {The Theory of
  Intermolecular Forces}}}\ (\bibinfo  {publisher} {Oxford University Press},\
  \bibinfo {address} {Oxford},\ \bibinfo {year} {2013})\BibitemShut {NoStop}%
\bibitem [{\citenamefont {Kaplan}(2006)}]{KaplanBook}%
  \BibitemOpen
  \bibfield  {author} {\bibinfo {author} {\bibfnamefont {I.~G.}\ \bibnamefont
  {Kaplan}},\ }\href@noop {} {\emph {\bibinfo {title} {Intermolecular
  Interactions: Physical Picture, Computational Methods and Model
  Potentials}}}\ (\bibinfo  {publisher} {J. Wiley and Sons},\ \bibinfo
  {address} {New York},\ \bibinfo {year} {2006})\BibitemShut {NoStop}%
\bibitem [{\citenamefont {Bellert}\ and\ \citenamefont
  {Breckenridge}(2002)}]{Breckenridge2002}%
  \BibitemOpen
  \bibfield  {author} {\bibinfo {author} {\bibfnamefont {D.}~\bibnamefont
  {Bellert}}\ and\ \bibinfo {author} {\bibfnamefont {W.~H.}\ \bibnamefont
  {Breckenridge}},\ }\href {\doibase 10.1021/cr980090e} {\bibfield  {journal}
  {\bibinfo  {journal} {Chem. Rev.}\ }\textbf {\bibinfo {volume} {102}},\
  \bibinfo {pages} {1595} (\bibinfo {year} {2002})}\BibitemShut {NoStop}%
\bibitem [{\citenamefont {Wright}\ and\ \citenamefont
  {Breckenridge}(2010)}]{WrightBreckenridge2010}%
  \BibitemOpen
  \bibfield  {author} {\bibinfo {author} {\bibfnamefont {T.~G.}\ \bibnamefont
  {Wright}}\ and\ \bibinfo {author} {\bibfnamefont {W.~H.}\ \bibnamefont
  {Breckenridge}},\ }\href {\doibase 10.1021/jp9091927} {\bibfield  {journal}
  {\bibinfo  {journal} {J. Phys. Chem. A}\ }\textbf {\bibinfo {volume} {114}},\
  \bibinfo {pages} {3182} (\bibinfo {year} {2010})}\BibitemShut {NoStop}%
\bibitem [{\citenamefont {Laatiaoui}(2019)}]{LRC2019}%
  \BibitemOpen
  \bibfield  {author} {\bibinfo {author} {\bibfnamefont {M.}~\bibnamefont
  {Laatiaoui}},\ }in\ \href
  {https://nyx.physics.mcgill.ca/event/149/book-of-abstracts.pdf} {\emph
  {\bibinfo {booktitle} {The 13th International Conference on Stopping and
  Manipulation of Ions and related topics (SMI-2019), Book of Abstracts}}}\
  (\bibinfo {year} {2019})\ p.~\bibinfo {pages} {8}\BibitemShut {NoStop}%
\bibitem [{\citenamefont {Viehland}(1983)}]{Viehland1983}%
  \BibitemOpen
  \bibfield  {author} {\bibinfo {author} {\bibfnamefont {L.~A.}\ \bibnamefont
  {Viehland}},\ }\href {\doibase 10.1016/0301-0104(83)85114-3} {\bibfield
  {journal} {\bibinfo  {journal} {Chem. Phys.}\ }\textbf {\bibinfo {volume}
  {78}},\ \bibinfo {pages} {279} (\bibinfo {year} {1983})}\BibitemShut
  {NoStop}%
\bibitem [{\citenamefont {Viehland}\ \emph {et~al.}(1976)\citenamefont
  {Viehland}, \citenamefont {Harrington},\ and\ \citenamefont
  {Mason}}]{Viehland1976}%
  \BibitemOpen
  \bibfield  {author} {\bibinfo {author} {\bibfnamefont {L.~A.}\ \bibnamefont
  {Viehland}}, \bibinfo {author} {\bibfnamefont {M.~M.}\ \bibnamefont
  {Harrington}}, \ and\ \bibinfo {author} {\bibfnamefont {E.~A.}\ \bibnamefont
  {Mason}},\ }\href {\doibase 10.1016/S0301-0104(76)80007-9} {\bibfield
  {journal} {\bibinfo  {journal} {Chem. Phys.}\ }\textbf {\bibinfo {volume}
  {17}},\ \bibinfo {pages} {433} (\bibinfo {year} {1976})}\BibitemShut
  {NoStop}%
\bibitem [{\citenamefont {Viehland}(1994)}]{ViehlandGC1994}%
  \BibitemOpen
  \bibfield  {author} {\bibinfo {author} {\bibfnamefont {L.~A.}\ \bibnamefont
  {Viehland}},\ }\href {\doibase 10.1016/0301-0104(93)E0337-U} {\bibfield
  {journal} {\bibinfo  {journal} {Chem. Phys.}\ }\textbf {\bibinfo {volume}
  {179}},\ \bibinfo {pages} {71} (\bibinfo {year} {1994})}\BibitemShut
  {NoStop}%
\bibitem [{\citenamefont {Viehland}(2012)}]{Viehland2012}%
  \BibitemOpen
  \bibfield  {author} {\bibinfo {author} {\bibfnamefont {L.~A.}\ \bibnamefont
  {Viehland}},\ }\href {\doibase 10.1007/s12127-011-0079-4} {\bibfield
  {journal} {\bibinfo  {journal} {Int. J. Ion Mobility Spectrom.}\ }\textbf
  {\bibinfo {volume} {15}},\ \bibinfo {pages} {21} (\bibinfo {year}
  {2012})}\BibitemShut {NoStop}%
\bibitem [{\citenamefont {Viehland}\ \emph {et~al.}(2017)\citenamefont
  {Viehland}, \citenamefont {Skaist}, \citenamefont {Adhikari},\ and\
  \citenamefont {Siems}}]{Viehland2017}%
  \BibitemOpen
  \bibfield  {author} {\bibinfo {author} {\bibfnamefont {L.~A.}\ \bibnamefont
  {Viehland}}, \bibinfo {author} {\bibfnamefont {T.}~\bibnamefont {Skaist}},
  \bibinfo {author} {\bibfnamefont {C.}~\bibnamefont {Adhikari}}, \ and\
  \bibinfo {author} {\bibfnamefont {W.~F.}\ \bibnamefont {Siems}},\ }\href
  {\doibase 10.1007/s12127-016-0212-5} {\bibfield  {journal} {\bibinfo
  {journal} {Int. J. Ion Mobility Spectrom.}\ }\textbf {\bibinfo {volume}
  {20}},\ \bibinfo {pages} {1} (\bibinfo {year} {2017})}\BibitemShut {NoStop}%
\bibitem [{\citenamefont {Viehland}(2020)}]{LXCAT}%
  \BibitemOpen
  \bibfield  {author} {\bibinfo {author} {\bibfnamefont {L.~A.}\ \bibnamefont
  {Viehland}},\ }\href {http://www.lxcat.net/Viehland} {\  (\bibinfo {year}
  {2009-2020})}\BibitemShut {NoStop}%
\bibitem [{\citenamefont {Hirschfelder}\ \emph {et~al.}(1954)\citenamefont
  {Hirschfelder}, \citenamefont {Curtiss},\ and\ \citenamefont
  {Bird}}]{Hirschfelder1954}%
  \BibitemOpen
  \bibfield  {author} {\bibinfo {author} {\bibfnamefont {J.~O.}\ \bibnamefont
  {Hirschfelder}}, \bibinfo {author} {\bibfnamefont {C.~F.}\ \bibnamefont
  {Curtiss}}, \ and\ \bibinfo {author} {\bibfnamefont {R.~B.}\ \bibnamefont
  {Bird}},\ }\href@noop {} {\emph {\bibinfo {title} {Molecular Theory of Gases
  and Liquids}}}\ (\bibinfo  {publisher} {J. Wiley and Sons},\ \bibinfo
  {address} {New York},\ \bibinfo {year} {1954})\BibitemShut {NoStop}%
\bibitem [{\citenamefont {Viehland}\ and\ \citenamefont
  {Chang}(2010)}]{PC2010}%
  \BibitemOpen
  \bibfield  {author} {\bibinfo {author} {\bibfnamefont {L.~A.}\ \bibnamefont
  {Viehland}}\ and\ \bibinfo {author} {\bibfnamefont {Y.}~\bibnamefont
  {Chang}},\ }\href {\doibase 10.1016/j.cpc.2010.06.008} {\bibfield  {journal}
  {\bibinfo  {journal} {Comput. Phys. Commun.}\ }\textbf {\bibinfo {volume}
  {181}},\ \bibinfo {pages} {1687} (\bibinfo {year} {2010})}\BibitemShut
  {NoStop}%
\bibitem [{\citenamefont {Aquilanti}\ and\ \citenamefont
  {Grossi}(1980)}]{AquilantiGrossi1980}%
  \BibitemOpen
  \bibfield  {author} {\bibinfo {author} {\bibfnamefont {V.}~\bibnamefont
  {Aquilanti}}\ and\ \bibinfo {author} {\bibfnamefont {G.}~\bibnamefont
  {Grossi}},\ }\href {\doibase 10.1063/1.440270} {\bibfield  {journal}
  {\bibinfo  {journal} {J. Chem. Phys.}\ }\textbf {\bibinfo {volume} {73}},\
  \bibinfo {pages} {1165} (\bibinfo {year} {1980})}\BibitemShut {NoStop}%
\bibitem [{\citenamefont {Krems}\ \emph {et~al.}(2004)\citenamefont {Krems},
  \citenamefont {Groenenboom},\ and\ \citenamefont {Dalgarno}}]{Krems2004}%
  \BibitemOpen
  \bibfield  {author} {\bibinfo {author} {\bibfnamefont {R.~V.}\ \bibnamefont
  {Krems}}, \bibinfo {author} {\bibfnamefont {G.~C.}\ \bibnamefont
  {Groenenboom}}, \ and\ \bibinfo {author} {\bibfnamefont {A.}~\bibnamefont
  {Dalgarno}},\ }\href {\doibase 10.1021/jp0488416} {\bibfield  {journal}
  {\bibinfo  {journal} {J. Phys. Chem. A}\ }\textbf {\bibinfo {volume} {108}},\
  \bibinfo {pages} {8941} (\bibinfo {year} {2004})}\BibitemShut {NoStop}%
\bibitem [{\citenamefont {Aquilanti}\ and\ \citenamefont
  {Vecchiocattivi}(1989)}]{Aquilanti1989}%
  \BibitemOpen
  \bibfield  {author} {\bibinfo {author} {\bibfnamefont {V.}~\bibnamefont
  {Aquilanti}}\ and\ \bibinfo {author} {\bibfnamefont {F.}~\bibnamefont
  {Vecchiocattivi}},\ }\href {\doibase 10.1016/0009-2614(89)87091-5} {\bibfield
   {journal} {\bibinfo  {journal} {Chem. Phys. Lett.}\ }\textbf {\bibinfo
  {volume} {156}},\ \bibinfo {pages} {109} (\bibinfo {year}
  {1989})}\BibitemShut {NoStop}%
\bibitem [{\citenamefont {Buchachenko}\ and\ \citenamefont
  {Viehland}(2019)}]{IJMS2}%
  \BibitemOpen
  \bibfield  {author} {\bibinfo {author} {\bibfnamefont {A.~A.}\ \bibnamefont
  {Buchachenko}}\ and\ \bibinfo {author} {\bibfnamefont {L.~A.}\ \bibnamefont
  {Viehland}},\ }\href {\doibase 10.1016/j.ijms.2019.06.005} {\bibfield
  {journal} {\bibinfo  {journal} {Int. J. Mass Spectrom.}\ }\textbf {\bibinfo
  {volume} {443}},\ \bibinfo {pages} {86} (\bibinfo {year} {2019})}\BibitemShut
  {NoStop}%
\bibitem [{\citenamefont {Laatiaoui}\ \emph {et~al.}(2012)\citenamefont
  {Laatiaoui}, \citenamefont {Backe}, \citenamefont {Habs}, \citenamefont
  {Kunz}, \citenamefont {Lauth},\ and\ \citenamefont {Sewtz}}]{Mustapha2012}%
  \BibitemOpen
  \bibfield  {author} {\bibinfo {author} {\bibfnamefont {M.}~\bibnamefont
  {Laatiaoui}}, \bibinfo {author} {\bibfnamefont {H.}~\bibnamefont {Backe}},
  \bibinfo {author} {\bibfnamefont {D.}~\bibnamefont {Habs}}, \bibinfo {author}
  {\bibfnamefont {P.}~\bibnamefont {Kunz}}, \bibinfo {author} {\bibfnamefont
  {W.}~\bibnamefont {Lauth}}, \ and\ \bibinfo {author} {\bibfnamefont
  {M.}~\bibnamefont {Sewtz}},\ }\href {\doibase 10.1140/epjd/e2012-30221-3}
  {\bibfield  {journal} {\bibinfo  {journal} {Eur. Phys. J. D}\ }\textbf
  {\bibinfo {volume} {66}},\ \bibinfo {pages} {232} (\bibinfo {year}
  {2012})}\BibitemShut {NoStop}%
\bibitem [{\citenamefont {Lee}\ and\ \citenamefont {Wright}(2013)}]{TimEd}%
  \BibitemOpen
  \bibfield  {author} {\bibinfo {author} {\bibfnamefont {E.~P.~F.}\
  \bibnamefont {Lee}}\ and\ \bibinfo {author} {\bibfnamefont {T.~G.}\
  \bibnamefont {Wright}},\ }\href@noop {} {\  (\bibinfo {year}
  {2011-2013})}\BibitemShut {NoStop}%
\bibitem [{\citenamefont {Buchachenko}\ and\ \citenamefont
  {Viehland}(2014)}]{JCP52}%
  \BibitemOpen
  \bibfield  {author} {\bibinfo {author} {\bibfnamefont {A.~A.}\ \bibnamefont
  {Buchachenko}}\ and\ \bibinfo {author} {\bibfnamefont {L.~A.}\ \bibnamefont
  {Viehland}},\ }\href {\doibase 10.1063/1.4868102} {\bibfield  {journal}
  {\bibinfo  {journal} {J. Chem. Phys.}\ }\textbf {\bibinfo {volume} {140}},\
  \bibinfo {pages} {114309} (\bibinfo {year} {2014})}\BibitemShut {NoStop}%
\bibitem [{\citenamefont {Manard}\ and\ \citenamefont
  {Kemper}(2017{\natexlab{a}})}]{Manard2017}%
  \BibitemOpen
  \bibfield  {author} {\bibinfo {author} {\bibfnamefont {M.~J.}\ \bibnamefont
  {Manard}}\ and\ \bibinfo {author} {\bibfnamefont {P.~R.}\ \bibnamefont
  {Kemper}},\ }\href {\doibase 10.1016/j.ijms.2016.11.015} {\bibfield
  {journal} {\bibinfo  {journal} {Int. J. Mass Spectrom.}\ }\textbf {\bibinfo
  {volume} {412}},\ \bibinfo {pages} {14} (\bibinfo {year}
  {2017}{\natexlab{a}})}\BibitemShut {NoStop}%
\bibitem [{\citenamefont {Manard}\ and\ \citenamefont
  {Kemper}(2017{\natexlab{b}})}]{Manard2017all}%
  \BibitemOpen
  \bibfield  {author} {\bibinfo {author} {\bibfnamefont {M.~J.}\ \bibnamefont
  {Manard}}\ and\ \bibinfo {author} {\bibfnamefont {P.~R.}\ \bibnamefont
  {Kemper}},\ }\href {\doibase 10.1016/j.ijms.2017.10.010} {\bibfield
  {journal} {\bibinfo  {journal} {Int. J. Mass Spectrom.}\ }\textbf {\bibinfo
  {volume} {423}},\ \bibinfo {pages} {54} (\bibinfo {year}
  {2017}{\natexlab{b}})}\BibitemShut {NoStop}%
\bibitem [{\citenamefont {Dolg}\ \emph {et~al.}(1989)\citenamefont {Dolg},
  \citenamefont {Stoll},\ and\ \citenamefont {Preuss}}]{ECP28MWB}%
  \BibitemOpen
  \bibfield  {author} {\bibinfo {author} {\bibfnamefont {M.}~\bibnamefont
  {Dolg}}, \bibinfo {author} {\bibfnamefont {H.}~\bibnamefont {Stoll}}, \ and\
  \bibinfo {author} {\bibfnamefont {H.}~\bibnamefont {Preuss}},\ }\href
  {\doibase 10.1063/1.456066} {\bibfield  {journal} {\bibinfo  {journal} {J.
  Chem. Phys.}\ }\textbf {\bibinfo {volume} {90}},\ \bibinfo {pages} {1730}
  (\bibinfo {year} {1989})}\BibitemShut {NoStop}%
\bibitem [{\citenamefont {Cao}\ and\ \citenamefont {Dolg}(2001)}]{LaANO}%
  \BibitemOpen
  \bibfield  {author} {\bibinfo {author} {\bibfnamefont {X.}~\bibnamefont
  {Cao}}\ and\ \bibinfo {author} {\bibfnamefont {M.}~\bibnamefont {Dolg}},\
  }\href {\doibase 10.1063/1.1406535} {\bibfield  {journal} {\bibinfo
  {journal} {J. Chem. Phys.}\ }\textbf {\bibinfo {volume} {115}},\ \bibinfo
  {pages} {7348} (\bibinfo {year} {2001})}\BibitemShut {NoStop}%
\bibitem [{\citenamefont {Buchachenko}\ \emph {et~al.}(2007)\citenamefont
  {Buchachenko}, \citenamefont {Cha{\l}asi{\'n}ski},\ and\ \citenamefont
  {Szcz\c{e}{\'s}niak}}]{SC1}%
  \BibitemOpen
  \bibfield  {author} {\bibinfo {author} {\bibfnamefont {A.~A.}\ \bibnamefont
  {Buchachenko}}, \bibinfo {author} {\bibfnamefont {G.}~\bibnamefont
  {Cha{\l}asi{\'n}ski}}, \ and\ \bibinfo {author} {\bibfnamefont {M.~M.}\
  \bibnamefont {Szcz\c{e}{\'s}niak}},\ }\href {\doibase
  10.1007/s11224-007-9243-1} {\bibfield  {journal} {\bibinfo  {journal}
  {Struct. Chem.}\ }\textbf {\bibinfo {volume} {18}},\ \bibinfo {pages} {769}
  (\bibinfo {year} {2007})}\BibitemShut {NoStop}%
\bibitem [{\citenamefont {Cao}\ and\ \citenamefont {Dolg}(2002)}]{LaSegm}%
  \BibitemOpen
  \bibfield  {author} {\bibinfo {author} {\bibfnamefont {X.}~\bibnamefont
  {Cao}}\ and\ \bibinfo {author} {\bibfnamefont {M.}~\bibnamefont {Dolg}},\
  }\href {\doibase 10.1016/S0166-1280(01)00751-5} {\bibfield  {journal}
  {\bibinfo  {journal} {J. Mol. Struct. (THEOCHEM)}\ }\textbf {\bibinfo
  {volume} {581}},\ \bibinfo {pages} {139} (\bibinfo {year}
  {2002})}\BibitemShut {NoStop}%
\bibitem [{\citenamefont {Woon}\ and\ \citenamefont {Dunning}(1994)}]{AVQZHe}%
  \BibitemOpen
  \bibfield  {author} {\bibinfo {author} {\bibfnamefont {D.~E.}\ \bibnamefont
  {Woon}}\ and\ \bibinfo {author} {\bibfnamefont {T.~H.}\ \bibnamefont
  {Dunning}},\ }\href {\doibase 10.1063/1.466439} {\bibfield  {journal}
  {\bibinfo  {journal} {J. Chem. Phys.}\ }\textbf {\bibinfo {volume} {100}},\
  \bibinfo {pages} {2975} (\bibinfo {year} {1994})}\BibitemShut {NoStop}%
\bibitem [{\citenamefont {Cybulski}\ and\ \citenamefont
  {Toczy{\l}owski}(1999)}]{Bf}%
  \BibitemOpen
  \bibfield  {author} {\bibinfo {author} {\bibfnamefont {S.~M.}\ \bibnamefont
  {Cybulski}}\ and\ \bibinfo {author} {\bibfnamefont {R.~R.}\ \bibnamefont
  {Toczy{\l}owski}},\ }\href {\doibase 10.1063/1.480430} {\bibfield  {journal}
  {\bibinfo  {journal} {J. Chem. Phys.}\ }\textbf {\bibinfo {volume} {111}},\
  \bibinfo {pages} {10520} (\bibinfo {year} {1999})}\BibitemShut {NoStop}%
\bibitem [{\citenamefont {Boys}\ and\ \citenamefont {Bernardi}(1970)}]{BoysB}%
  \BibitemOpen
  \bibfield  {author} {\bibinfo {author} {\bibfnamefont {S.~F.}\ \bibnamefont
  {Boys}}\ and\ \bibinfo {author} {\bibfnamefont {F.}~\bibnamefont
  {Bernardi}},\ }\href {\doibase 10.1080/00268977000101561} {\bibfield
  {journal} {\bibinfo  {journal} {Mol. Phys.}\ }\textbf {\bibinfo {volume}
  {19}},\ \bibinfo {pages} {553} (\bibinfo {year} {1970})}\BibitemShut
  {NoStop}%
\bibitem [{\citenamefont {Werner}\ \emph {et~al.}(2015)\citenamefont {Werner},
  \citenamefont {Knowles}, \citenamefont {Knizia}, \citenamefont {Manby},
  \citenamefont {Sch{\"u}tz} \emph {et~al.}}]{MOLPRO}%
  \BibitemOpen
  \bibfield  {author} {\bibinfo {author} {\bibfnamefont {H.-J.}\ \bibnamefont
  {Werner}}, \bibinfo {author} {\bibfnamefont {P.~J.}\ \bibnamefont {Knowles}},
  \bibinfo {author} {\bibfnamefont {G.}~\bibnamefont {Knizia}}, \bibinfo
  {author} {\bibfnamefont {F.~R.}\ \bibnamefont {Manby}}, \bibinfo {author}
  {\bibfnamefont {M.}~\bibnamefont {Sch{\"u}tz}},  \emph {et~al.},\ }\href
  {http://www.molpro.net/} {\  (\bibinfo {year} {2015})}\BibitemShut {NoStop}%
\bibitem [{\citenamefont {Backe}\ \emph {et~al.}(2005)\citenamefont {Backe},
  \citenamefont {Dretzke}, \citenamefont {Horn}, \citenamefont {Kolb},
  \citenamefont {Lauth}, \citenamefont {Repnow}, \citenamefont {Sewtz},\ and\
  \citenamefont {Trautmann}}]{Backe2005}%
  \BibitemOpen
  \bibfield  {author} {\bibinfo {author} {\bibfnamefont {H.}~\bibnamefont
  {Backe}}, \bibinfo {author} {\bibfnamefont {A.}~\bibnamefont {Dretzke}},
  \bibinfo {author} {\bibfnamefont {R.}~\bibnamefont {Horn}}, \bibinfo {author}
  {\bibfnamefont {T.}~\bibnamefont {Kolb}}, \bibinfo {author} {\bibfnamefont
  {W.}~\bibnamefont {Lauth}}, \bibinfo {author} {\bibfnamefont
  {R.}~\bibnamefont {Repnow}}, \bibinfo {author} {\bibfnamefont
  {M.}~\bibnamefont {Sewtz}}, \ and\ \bibinfo {author} {\bibfnamefont
  {N.}~\bibnamefont {Trautmann}},\ }\href {\doibase 10.1007/s10751-005-9210-4}
  {\bibfield  {journal} {\bibinfo  {journal} {Hyperf. Int.}\ }\textbf {\bibinfo
  {volume} {162}},\ \bibinfo {pages} {77} (\bibinfo {year} {2005})}\BibitemShut
  {NoStop}%
\bibitem [{\citenamefont {Indelicato}\ \emph {et~al.}(2007)\citenamefont
  {Indelicato}, \citenamefont {Santos}, \citenamefont {Boucard},\ and\
  \citenamefont {Desclaux}}]{Indelicato2007}%
  \BibitemOpen
  \bibfield  {author} {\bibinfo {author} {\bibfnamefont {P.}~\bibnamefont
  {Indelicato}}, \bibinfo {author} {\bibfnamefont {J.~P.}\ \bibnamefont
  {Santos}}, \bibinfo {author} {\bibfnamefont {S.}~\bibnamefont {Boucard}}, \
  and\ \bibinfo {author} {\bibfnamefont {J.-P.}\ \bibnamefont {Desclaux}},\
  }\href {\doibase 10.1140/epjd/e2007-00229-y} {\bibfield  {journal} {\bibinfo
  {journal} {Eur. Phys. J. D}\ }\textbf {\bibinfo {volume} {45}},\ \bibinfo
  {pages} {155} (\bibinfo {year} {2007})}\BibitemShut {NoStop}%
\bibitem [{\citenamefont {Johnsen}\ and\ \citenamefont
  {Biondi}(1972)}]{Biondi1972}%
  \BibitemOpen
  \bibfield  {author} {\bibinfo {author} {\bibfnamefont {R.}~\bibnamefont
  {Johnsen}}\ and\ \bibinfo {author} {\bibfnamefont {M.~A.}\ \bibnamefont
  {Biondi}},\ }\href {\doibase 10.1063/1.1678220} {\bibfield  {journal}
  {\bibinfo  {journal} {J. Chem. Phys.}\ }\textbf {\bibinfo {volume} {57}},\
  \bibinfo {pages} {5292} (\bibinfo {year} {1972})}\BibitemShut {NoStop}%
\bibitem [{\citenamefont {Ellis}\ \emph {et~al.}(1976)\citenamefont {Ellis},
  \citenamefont {Pai}, \citenamefont {McDaniel}, \citenamefont {Mason},\ and\
  \citenamefont {Viehland}}]{EllisADNDT}%
  \BibitemOpen
  \bibfield  {author} {\bibinfo {author} {\bibfnamefont {H.~W.}\ \bibnamefont
  {Ellis}}, \bibinfo {author} {\bibfnamefont {R.~Y.}\ \bibnamefont {Pai}},
  \bibinfo {author} {\bibfnamefont {E.~W.}\ \bibnamefont {McDaniel}}, \bibinfo
  {author} {\bibfnamefont {E.~A.}\ \bibnamefont {Mason}}, \ and\ \bibinfo
  {author} {\bibfnamefont {L.~A.}\ \bibnamefont {Viehland}},\ }\href {\doibase
  10.1016/0092-640X(76)90001-2} {\bibfield  {journal} {\bibinfo  {journal} {At.
  Data Nucl. Data Tables}\ }\textbf {\bibinfo {volume} {17}},\ \bibinfo {pages}
  {177} (\bibinfo {year} {1976})}\BibitemShut {NoStop}%
\bibitem [{\citenamefont {Lee}\ \emph {et~al.}(2011)\citenamefont {Lee},
  \citenamefont {Viehland}, \citenamefont {Johnsen}, \citenamefont
  {Breckenridge},\ and\ \citenamefont {Wright}}]{TimUplus}%
  \BibitemOpen
  \bibfield  {author} {\bibinfo {author} {\bibfnamefont {E.~P.~F.}\
  \bibnamefont {Lee}}, \bibinfo {author} {\bibfnamefont {L.~A.}\ \bibnamefont
  {Viehland}}, \bibinfo {author} {\bibfnamefont {R.}~\bibnamefont {Johnsen}},
  \bibinfo {author} {\bibfnamefont {W.~H.}\ \bibnamefont {Breckenridge}}, \
  and\ \bibinfo {author} {\bibfnamefont {T.~G.}\ \bibnamefont {Wright}},\
  }\href {\doibase 10.1021/jp2076879} {\bibfield  {journal} {\bibinfo
  {journal} {J. Phys. Chem. A}\ }\textbf {\bibinfo {volume} {115}},\ \bibinfo
  {pages} {12126} (\bibinfo {year} {2011})}\BibitemShut {NoStop}%
\bibitem [{\citenamefont {K{\"u}chle}\ \emph {et~al.}(1994)\citenamefont
  {K{\"u}chle}, \citenamefont {Dolg}, \citenamefont {Stoll},\ and\
  \citenamefont {Preuss}}]{Kuchle1994}%
  \BibitemOpen
  \bibfield  {author} {\bibinfo {author} {\bibfnamefont {W.}~\bibnamefont
  {K{\"u}chle}}, \bibinfo {author} {\bibfnamefont {M.}~\bibnamefont {Dolg}},
  \bibinfo {author} {\bibfnamefont {H.}~\bibnamefont {Stoll}}, \ and\ \bibinfo
  {author} {\bibfnamefont {H.}~\bibnamefont {Preuss}},\ }\href {\doibase
  10.1063/1.466847} {\bibfield  {journal} {\bibinfo  {journal} {J. Chem.
  Phys.}\ }\textbf {\bibinfo {volume} {100}},\ \bibinfo {pages} {7535}
  (\bibinfo {year} {1994})}\BibitemShut {NoStop}%
\bibitem [{\citenamefont {Cao}\ and\ \citenamefont {Dolg}(2004)}]{AcSegm}%
  \BibitemOpen
  \bibfield  {author} {\bibinfo {author} {\bibfnamefont {X.}~\bibnamefont
  {Cao}}\ and\ \bibinfo {author} {\bibfnamefont {M.}~\bibnamefont {Dolg}},\
  }\href {\doibase 10.1016/j.theochem.2003.12.015} {\bibfield  {journal}
  {\bibinfo  {journal} {J. Molec. Struct. (THEOCHEM)}\ }\textbf {\bibinfo
  {volume} {673}},\ \bibinfo {pages} {203} (\bibinfo {year}
  {2004})}\BibitemShut {NoStop}%
\bibitem [{\citenamefont {Moritz}\ \emph {et~al.}(2007)\citenamefont {Moritz},
  \citenamefont {Cao},\ and\ \citenamefont {Dolg}}]{LCECP}%
  \BibitemOpen
  \bibfield  {author} {\bibinfo {author} {\bibfnamefont {A.}~\bibnamefont
  {Moritz}}, \bibinfo {author} {\bibfnamefont {X.}~\bibnamefont {Cao}}, \ and\
  \bibinfo {author} {\bibfnamefont {M.}~\bibnamefont {Dolg}},\ }\href {\doibase
  10.1007/s00214-006-0180-7} {\bibfield  {journal} {\bibinfo  {journal} {Theor.
  Chem. Acc.}\ }\textbf {\bibinfo {volume} {117}},\ \bibinfo {pages} {473}
  (\bibinfo {year} {2007})}\BibitemShut {NoStop}%
\bibitem [{\citenamefont {Evangelista}(2018)}]{MRCC}%
  \BibitemOpen
  \bibfield  {author} {\bibinfo {author} {\bibfnamefont {F.~A.}\ \bibnamefont
  {Evangelista}},\ }\href {\doibase 10.1063/1.5039496} {\bibfield  {journal}
  {\bibinfo  {journal} {J. Chem. Phys.}\ }\textbf {\bibinfo {volume} {149}},\
  \bibinfo {pages} {030901} (\bibinfo {year} {2018})}\BibitemShut {NoStop}%
\bibitem [{\citenamefont {Viehland}(2016)}]{Viehland2016}%
  \BibitemOpen
  \bibfield  {author} {\bibinfo {author} {\bibfnamefont {L.~A.}\ \bibnamefont
  {Viehland}},\ }\href {\doibase 10.1007/s12127-016-0192-5} {\bibfield
  {journal} {\bibinfo  {journal} {Int. J. Ion Mobility Spectrom.}\ }\textbf
  {\bibinfo {volume} {19}},\ \bibinfo {pages} {1} (\bibinfo {year}
  {2016})}\BibitemShut {NoStop}%
\end{thebibliography}%

%%% Make sure to upload the bib file along with the tex file and PDF
%%% Please see the test.bib file for some examples of references

%%% If you are submitting a figure with subfigures please combine these into one image file with part labels integrated.
%%% If you don't add the figures in the LaTeX files, please upload them when submitting the article.
%%% Frontiers will add the figures at the end of the provisional pdf automatically
%%% The use of LaTeX coding to draw Diagrams/Figures/Structures should be avoided. They should be external callouts including graphics.

\end{document}